\documentclass[twocolumn]{aastex62}
\usepackage{amsmath}
\received{}
\revised{}

\begin{document}

\title{ALMA Observations of Atomic Carbon [\ion{C}{1}] \(\left({^3}\mathrm{P}_1\rightarrow{^3}\mathrm{P}_0\right)\) and Low-\(J\) CO Lines in the Starburst Galaxy NGC 1808}


\correspondingauthor{Dragan Salak}
\email{salak.dragan.fm@u.tsukuba.ac.jp}

\author[0000-0002-3848-1757]{Dragan Salak}
\affil{Tomonaga Center for the History of the Universe, University of Tsukuba, 
1-1-1 Tennodai, Tsukuba,
Ibaraki 305-8571, Japan}

\author[0000-0002-5461-6359]{Naomasa Nakai}
\affil{School of Science and Technology, Kwansei Gakuin University, 
2-1 Gakuen, Sanda,
Hyogo 669-1337, Japan}
\affil{Faculty of Pure and Applied Sciences, University of Tsukuba, 
1-1-1 Tennodai, Tsukuba,
Ibaraki 305-8571, Japan}

\author{Masumichi Seta}
\affil{School of Science and Technology, Kwansei Gakuin University, 
2-1 Gakuen, Sanda,
Hyogo 669-1337, Japan}

\author[0000-0002-7616-7427]{Yusuke Miyamoto}
\affiliation{National Astronomical Observatory of Japan,
2-21-1 Osawa, Mitaka,
Tokyo, 181-8588, Japan}

\begin{abstract}

We present [\ion{C}{1}] \(\left({^3}\mathrm{P}_1\rightarrow{^3}\mathrm{P}_0\right)\), \(^{12}\)CO, \(^{13}\)CO, and C\(^{18}\)O (\(J=2\rightarrow1\)) observations of the central region (radius 1 kpc) of the starburst galaxy NGC 1808 at 30--50 pc resolution conducted with Atacama Large Millimeter/submillimeter Array. Radiative transfer analysis of multiline data indicates warm (\(T_\mathrm{k}\sim40\mathrm{-}80\) K) and dense (\(n_\mathrm{H_2}\sim10^{3\mathrm{-}4}\) cm\(^{-3}\)) molecular gas with high column density of atomic carbon (\(N_\mathrm{CI}\sim3\times10^{18}\) cm\(^{-2}\)) in the circumnuclear disk (central 100 pc). The \ion{C}{1}/H\(_2\) abundance in the central 1 kpc is \(\sim3\mathrm{-}7\times10^{-5}\), consistent with the values in luminous infrared galaxies. The intensity ratios of [\ion{C}{1}]/CO(1--0) and [\ion{C}{1}]/CO(3--2), respectively, decrease and increase with radius in the central 1 kpc, whereas [\ion{C}{1}]/CO(2--1) is uniform within statistical errors. The result can be explained by excitation and optical depth effects, since the effective critical density of CO (2--1) is comparable to that of [\ion{C}{1}].  The distribution of [\ion{C}{1}] is similar to that of \(^{13}\)CO(2--1), and the ratios of [\ion{C}{1}] to \(^{13}\)CO(2--1) and C\(^{18}\)O(2--1) are uniform within \(\sim30\%\) in the central \(<400\) pc starburst disk. The results suggest that [\ion{C}{1}] \(\left({^3}\mathrm{P}_1\rightarrow{^3}\mathrm{P}_0\right)\) luminosity can be used as a CO-equivalent tracer of molecular gas mass, although caution is needed when applied in resolved starburst nuclei (e.g., circumnuclear disk), where the [\ion{C}{1}]/CO(1--0) luminosity ratio is enhanced due to high excitation and atomic carbon abundance. The [\ion{C}{1}]/CO(1--0) intensity ratio toward the base of the starburst-driven outflow is \(\lesssim0.15\), and the upper limits of the mass and kinetic energy of the atomic carbon outflow are \(\sim1\times10^4~M_\sun\) and \(\sim3\times10^{51}~\mathrm{erg}\), respectively.

\end{abstract}


\section{Introduction}\label{sec:intro}

Neutral atomic carbon (\ion{C}{1}) is a major constituent and plays an important role in the physics and chemistry of the cold interstellar medium (ISM) in galaxies. The \ion{C}{1} gas is observable in the fine-structure transitions [\ion{C}{1}] (\(2p^2:{^3}\mathrm{P}_1\rightarrow{^3}\mathrm{P}_0\)), hereafter [\ion{C}{1}] (1--0), and [\ion{C}{1}] ( \(2p^2:{^3}\mathrm{P}_2\rightarrow{^3}\mathrm{P}_1\)) in the submillimeter regime \citep{Phi80}. Owing to the high abundance of carbon atoms and low energies above the ground state for the fine-structure levels, the two lines of the triplet are important cooling channels in the neutral ISM, comparable to those of low-\(J\) CO lines (e.g., \citealt{Pen70}). The critical density of [\ion{C}{1}] (1--0) is \(n_\mathrm{cr}/\beta\sim1\times10^3\) cm\(^{-3}\), which is similar to that of CO (1--0), suggesting that both [\ion{C}{1}] and CO (1--0) are primary tracers of molecular gas.\footnote{The effective critical density can be defined as \(n_\mathrm{cr}\equiv \beta A_\mathrm{ul}/{C_\mathrm{ul}}\), where \(\beta\) is the photon escape probability, \(A_\mathrm{ul}\) is the Einstein coefficient for spontaneous emission, and \(C_\mathrm{ul}\) is the coefficient for collisional deexcitation.}

[\ion{C}{1}] (1--0) observations have been conducted extensively toward prominent Galactic clouds (e.g., \citealt{Kee85,Kee97,Tau95,Ike99,Ike02,Tat99,Oka05,Rol11,Shi13,Beu14}) and the Galactic center (e.g., \citealt{Ojh01,Mar04,Tan11}). These observations have shown that the [\ion{C}{1}] distribution is typically similar to that of low-\(J\) CO lines in Galactic clouds. [\ion{C}{1}] (1--0) emission has also been observed toward gas-rich nearby galaxies with single dish telescopes (e.g., \citealt{Sch93,Stu97,GP00,PG04a,Kam12,IRv15}), and recently at high resolution with Atacama Large Millimeter/submillimeter Array (ALMA) (e.g., \citealt{Kri16,Izu18,Miy18}). The line has also been used to probe the ISM conditions in objects at high redshift (e.g., \citealt{Bar97,Wei05,Dan11,Wal11,AZ13,Bot17,Pop17,Emo18,Val18,Nes19}). These works indicate that [\ion{C}{1}] is luminous in molecular-gas rich galaxies and that the properties of [\ion{C}{1}] are not significantly different between the objects at high redshift and those in the local Universe.

While high-resolution [\ion{C}{1}] observations of Galactic molecular clouds reveal the distribution of atomic carbon with respect to CO on sub-pc scale, which is essential to understand the structure of photodissociation regions \citep{MT93,Spa96,HT97,SvD97}, extragalactic observations provide insight into the environmental effects and physical conditions on the scale of entire galaxies. Studying [\ion{C}{1}] is particularly important because it has been considered as a valuable tracer of molecular gas mass in distant galaxies \citep{AZ13,Tom14,Glo15,Off15,Yan17}. The cosmic microwave background at high redshift (\(z\gtrsim2\)) significantly affects the observed flux at low frequencies, such as those of the low-\(J\) CO lines, whereas the effect is less prominent at the frequencies of [\ion{C}{1}] \citep{daC13,Zha16}. Whether or not it can be a substitute for CO, however, is a matter of ongoing debate (e.g., \citealt{IRv15,Val18,GOB19}).

To understand the origin of [\ion{C}{1}] emission and its relation with CO in different environments in gas-rich, star-forming galaxies, it is important to bridge the gap between the spatial scales of individual Galactic clouds and poorly-resolved distant galaxies. This can now be done by wide-field imaging of nearby galaxies at high resolution using ALMA. Since central regions produce much of the observed CO and [\ion{C}{1}] flux in starburst galaxies, they are ideal laboratories to construct templates for distant galaxies. Toward this goal, we have conducted comprehensive observations of the central region (radius 1 kpc) of the starburst galaxy NGC 1808 in [\ion{C}{1}] (1--0) and five CO lines, including \(^{13}\)CO and C\(^{18}\)O, at a resolution of 30-50 pc.

\begin{figure}
 \centering
  \includegraphics[width=0.45\textwidth]{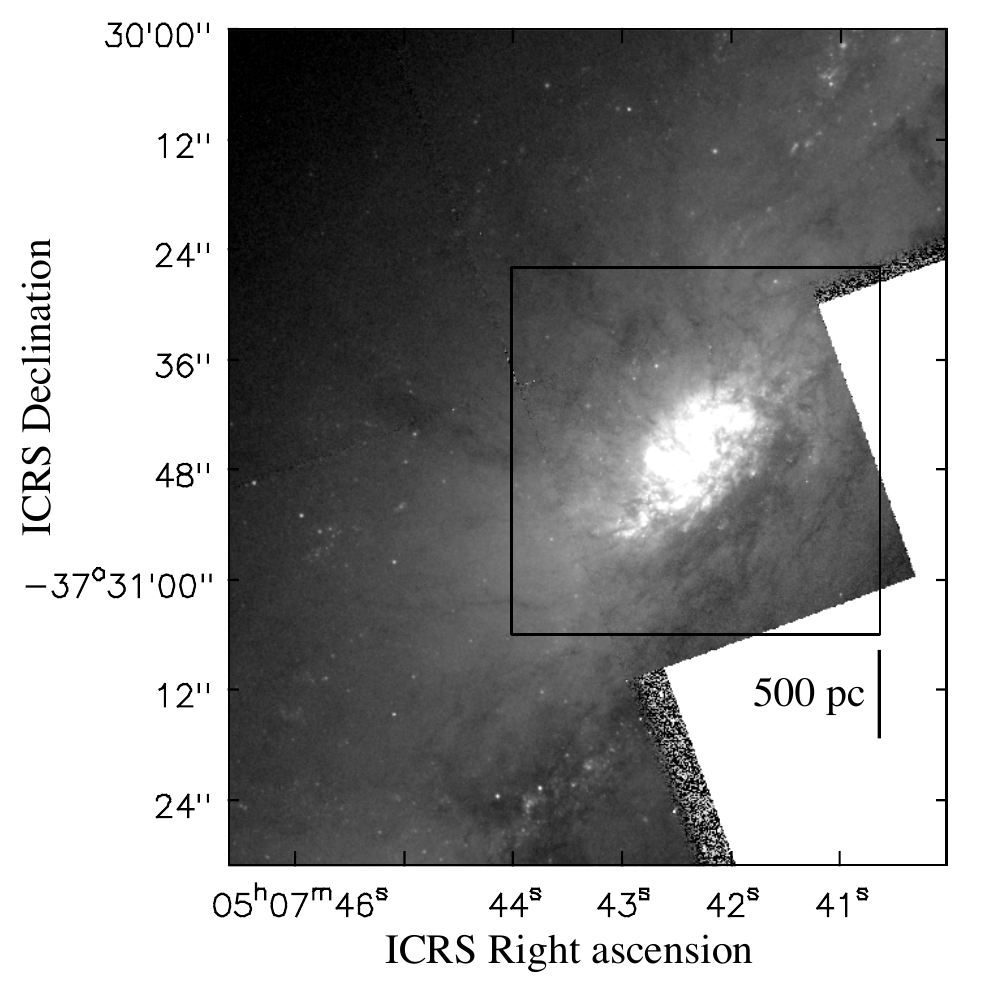}
 \caption{R (675W) image of NGC 1808 (Hubble Legacy Archive). The observed region is indicated by a square.\label{fig:gal}}
\end{figure}

The case-study object (Table \ref{tab:gal}, Figure \ref{fig:gal}) is a nearby barred galaxy with vigorous star formation in its central 1 kpc region, as revealed by the presence of \ion{H}{2} regions, young star clusters, and supernova remnants detected at various wavelengths (e.g., \citealt{Dah90,Sai90,Col94,Kot96,TG05,GA08,Bus17}). The galaxy has been classified as a starburst/Seyfert composite \citep{VV85}, but the activity appears to be dominated by star formation feedback, with a total star formation rate in the central 1 kpc of \(\sim4~M_\sun~\mathrm{yr}^{-1}\) \citep{Sal17}. On kpc scale, the most striking feature in optical images is the presence of polar dust lanes that appear to emerge from the central 1 kpc disk (e.g., \citealt{VV85,Phi93}; Figure \ref{fig:gal}). Observations of neutral gas suggest that the feature is a starburst-driven outflow from the central region \citep{Phi93}. So far, the neutral gas outflow has been identified kinematically in \ion{Na}{1} \citep{Phi93}, \ion{H}{1} \citep{Kor93}, and CO \citep{Sal16}. From morphology, there is evidence of extended emission from ionized gas tracers and polycyclic aromatic hydrocarbons \citep{SB10,McC13}. The molecular outflow is mostly within \(R\lesssim1\) kpc from the center and includes less than 10\% of the total molecular gas budget within that region \citep{Sal16,Sal17}. Since it has been suggested that \ion{C}{1} abundance may be elevated with respect to CO in starbursts and molecular outflows as a consequence of high cosmic ray flux \citep{PTV04,PBZ18,Bis17}, a key objective of this work was to search for [\ion{C}{1}] emission in the dust outflow and compare it with CO.

\begin{table}
\begin{center}
\caption{Basic Parameters of NGC 1808}\label{tab:gal}
\begin{tabular}{llc}
\tableline\tableline
Parameter & Value & Reference \\
\tableline
Morphological type & (R)SAB(s)a & (1) \\
\(\alpha_\mathrm{ICRS}\) & \(\mathrm{05^h07^m42\fs329}\) & (2) \\
\(\delta_\mathrm{ICRS}\) & \(-37\arcdeg30\arcmin45\farcs85\) & (2) \\
Distance & 10.8 Mpc (\(1\arcsec=52\) pc) & (3) \\
\(V_\mathrm{sys}\) (LSR) & 998 km s\(^{-1}\) (CND) & (4) \\
Position angle & 324\arcdeg & (4) \\
Inclination & 57\arcdeg & (5) \\
Central activity & starburst, Seyfert 2 & (6) \\
\tableline
\end{tabular}
\end{center}
\tablerefs{(1) \citet{deV91}, (2) \citet{Com19}, (3) \citet{Tul88}, (4) \citet{Sal16}, (5) \citet{Rei82}, (6) NED classification.}
\end{table}

\section{Observations}\label{sec:obs}

The ALMA observations were carried out in cycle 5 during 2017 and 2018 by the 12-m array, 7-m array (Atacama Compact Array; ACA), and total power (TP) antennas. Three tuning settings were implemented to cover the 220 GHz, 230 GHz, and 500 GHz frequency bands, where CO (2--1) and [\ion{C}{1}] (1--0) lines were the main target lines. Each setting included four basebands (each 1.875 GHz wide) that were observed simultaneously. To cover a rectangular field of \(40\arcsec\times40\arcsec\), designed to include the central starburst region, observations were done in mosaic mode. The coordinates of the mosaic center were \((\alpha,\delta)_\mathrm{J2000.0}=\mathrm{5^h7^m42^s.331,-37\arcdeg30'45\farcs88}\). The number of pointings in a mosaic varied with band and array, as shown in Table \ref{tab:obs}. The center positions of neighboring mosaic fields were separated from each other by one half the primary beam size. In order to image the entire galactic center region (radius 1 kpc corresponding to \(\sim20\arcsec\)) and fully recover the flux, it was essential to use ACA. The absolute flux uncertainty of the interferometer is 10\% in bands 6 and 8 and that of the TP is 15\% in band 8.

The acquired visibility data were reduced using the Common Astronomy Software Applications (CASA) package \citep{McM07} in the following order. First, the data were calibrated by a CASA pipeline. The calibrated data were then split into line and continuum data sets, and the line data were continuum-subtracted. These steps were performed separately for 12-m and ACA data. We then reconstructed images from the two visibility data sets using the task \emph{tclean} in CASA, yielding a single data product (12m+ACA). The CLEAN process was interactively performed by applying the Hogbom minor cycle algorithm, mosaic gridder, and Briggs weighting with the robust parameter of 0.5. The number of iterations was high enough so that the peak intensity in the residual images was comparable to the image rms. The spectral resolution in band 8 was \(\Delta v=9.5~\mathrm{km~s}^{-1}\) and the pixel size was set to be \(\sim1/5\) of the beam minor axis (\(0\farcs2\) in band 6 and \(0\farcs1\) in band 8). The image reconstruction was done in ``cube'' mode for the line data, and in ``mfs'' (multi-frequency synthesis) mode with one output image channel for the continuum data. The [\ion{C}{1}] (1--0) line data in band 8 were then combined with TP data using the CASA task \emph{feather}, which resulted in the final data cube presented in this paper. The final images were corrected for the primary beam response. 

The total flux in band 6 did not increase after feathering with TP, which may be because the maximum recoverable scale of ACA (\(\sim29\arcsec\)) was comparable to the starburst region. On the other hand, the maximum recoverable scale in band 8 was much smaller (\(\sim13\arcsec\)), and the total flux of the [\ion{C}{1}] (1--0) line across the reconstructed \(40\arcsec\times40\arcsec\) image increased from 2163 Jy km s\(^{-1}\) (12m+ACA) to 4948 Jy km s\(^{-1}\) (12m+ACA feathered with TP), yielding a recovery of as much as 56\% of the flux by combining the interferometer data with TP. We present the feathered [\ion{C}{1}] (1--0) image below.

The basic parameters of the observations are summarized in Table \ref{tab:obs}. The descriptions of the observations and data reduction of CO (1--0) and CO (3--2) data used in this paper, that also contain 12-m, ACA, and TP data, are given in \citet{Sal16,Sal17}. The total flux has been recovered in all presented data using the standard methods of image combining in CASA applied to 12-m, ACA, and TP data sets.

The velocity in all data is in radio definition with respect to the local standard of rest (LSR). Throughout the paper, CO refers to the \(^{12}\)CO isotopologue. The atomic carbon gas is denoted by \ion{C}{1}, whereas the fine-structure transitions are denoted by [\ion{C}{1}].

\begin{deluxetable*}{lcccccc}
\tablecaption{Observation Summary \label{tab:obs}}
\tablecolumns{6}
\tablewidth{0pt}
\tablehead{
\colhead{Parameter} &
\colhead{12m (B6)} &
\colhead{7m (B6)} &
\colhead{TP (B6)} &
\colhead{12m (B8)} & \colhead{7m (B8)} & \colhead{TP (B8)}
}
\startdata
Frequency band (GHz) & 220, 230 & 220, 230 & 220, 230 & 500 & 500 & 500 \\
Observation date & 2018 May 6 & 2017 Oct 19, & 2018 Jan 11, 15 & 2018 May 22 & 2017 Dec 25, & 2017 Dec 25, 26, \\
& & 31 & 21, 23 & & 2018 May 22 & 2018 May 12, 13 \\
Number of antennas & 43 & 10 & 3 & 43 & 10 & 3-4 \\
Number of pointings & 14 & 3, 5 & ... & 52 & 17 & ... \\
Baselines (meters) & 15-479 & 9-45 & ... & 15-314 & 9-45 & ... \\
Mosaic size & \(40\arcsec\times40\arcsec\) & \(40\arcsec\times40\arcsec\) & ... &  \(40\arcsec\times40\arcsec\) & \(40\arcsec\times40\arcsec\) & ... \\
Time on source (min.) & 7, 6 & 32, 25 & \(1.1\times2\), \(1.2\times2\) & 34 & \(44\times4\) & \(59\times7\) \\
Flux and bandpass & J0519-4546 & J0510+1800  & ... & J0423-0120 & J0854+2006 & Uranus \\
calibrator & & & & & J0423-0120 \\
 & & & & & J0510+1800 \\
 Phase calibrator & J0522-3627 & J0522-3627 & ... & J0522-3627 & J0522-3627 & ... \\
\enddata
\end{deluxetable*}

\begin{deluxetable*}{lccccccc}
\tablecaption{Data Summary \label{tab:res}}
\tablecolumns{7}
\tablewidth{0pt}
\tablehead{
\colhead{Transition} &
\colhead{Rest frequency\tablenotemark{a}} &
\colhead{\(E_\mathrm{u}/k\)\tablenotemark{b}} &
\colhead{Resolution\tablenotemark{c}} &
\colhead{Sensitivity} &
\colhead{\(F\)\tablenotemark{d}} &
\colhead{\(L'\)\tablenotemark{d}} & \\
\colhead{} & \colhead{(GHz)} & \colhead{(K)} & \colhead{(FWHM)} & \colhead{(mJy/beam)} & \colhead{(Jy km s\(^{-1}\))} & \colhead{(K km s\(^{-1}\) pc\(^2\))}
}
\startdata
CO (\(J=1\rightarrow0\))\tablenotemark{e} & 115.2712018 & 5.53 & \(2\farcs666\times1\farcs480\) & 4.2 & 2322 & \((6.60\pm0.66)\times10^8\) \\
C\(^{18}\)O (\(J=2\rightarrow1\)) & 219.5603541 & 15.8 & \(1\farcs286\times0\farcs967\) & 2.7 & 78.9 & \((6.18\pm0.62)\times10^6\) \\
HNCO (\(J_{K_a,K_c}=10_{0,10}\rightarrow9_{0,9}\)) & 219.7982740 & 58.0 & \(1\farcs286\times0\farcs967\) & 2.7 & 2.80 & \((2.19\pm0.22)\times10^5\) \\
\(^{13}\)CO (\(J=2\rightarrow1\)) & 220.3986842 & 15.9 & \(1\farcs262\times0\farcs975\) & 2.8 & 362 & \((2.82\pm0.28)\times10^7\) \\
CO (\(J=2\rightarrow1\)) & 230.5380000 & 16.6 & \(1\farcs280\times0\farcs943\) & 3.0 & 5424 & \((3.86\pm0.39)\times10^8\) \\
CS (\(J=5\rightarrow4\)) & 244.9355565 & 35.3 & \(1\farcs218\times0\farcs859\) & 3.6 & 10.4 & \((6.55\pm0.66)\times10^5\) \\
CO (\(J=3\rightarrow2\))\tablenotemark{e} & 345.7959899 & 33.2 & \(1\farcs047\times0\farcs568\) & 7.8 & 9908 & \((3.13\pm0.31)\times10^8\) \\
\text{[\ion{C}{1}]} (\({^3}\mathrm{P}_1\rightarrow{^3}\mathrm{P}_0\)) & 492.1606510 & 23.6 & \(0\farcs825\times0\farcs590\) & 20 & 4683 & \((7.30\pm1.01)\times10^7\) \\
\enddata
\tablenotetext{a}{Acquired from the Splatalogue data base: \url{http://www.cv.nrao.edu/php/splat/}.}
\tablenotetext{b}{The energy above ground state divided by the Boltzmann constant.}
\tablenotetext{c}{Full width half maximum (FWHM) of the major and minor beam axes (beam size).}
\tablenotetext{d}{Calculated from equation \ref{eq:lum} within an aperture of radius \(r=20\arcsec=1.04\) kpc, except HNCO, for which it is \(r=5\arcsec\) because emission is weak. The adopted absolute flux uncertainty is \(15\%\) for [\ion{C}{1}] and \(10\%\) for the rest. The luminosity can be converted to \(L_\sun\) units by \(L=3.20\times10^{-11}\nu_\mathrm{rest}^3L'\), where \(\nu_\mathrm{rest}\) is in GHz and \(L'\) is in K km s\(^{-1}\) pc\(^2\).}
\tablenotetext{e}{Data from \citet{Sal17}.}
\end{deluxetable*}

\section{Results}\label{sec:res}

In the following sections, we present the images of the detected emission lines in the central 1 kpc starburst region at 30--50 pc resolution. These are the first high resolution images of CO (2--1), \(^{13}\)CO (2--1), C\(^{18}\)O (2--1), CS (5--4), HNCO (10--9), and [\ion{C}{1}] (1--0). Only CO (2--1) was mapped earlier with a single dish telescope \citep{Aal94}. The basic properties of the line data are listed in Table \ref{tab:res}. The results for CO (2--1) and [\ion{C}{1}] (1--0) are presented in sections \ref{sec:cod} and \ref{sec:cid}, respectively, whereas those for the dense gas tracers \(^{13}\)CO (2--1), C\(^{18}\)O (2--1), CS (5--4), and HNCO (10--9) are presented in section \ref{sec:dgt}.

The line luminosity \(L'\) in Table \ref{tab:res} is calculated from

\begin{eqnarray}\label{eq:lum}
\left(\frac{L'}{\mathrm{K~km~s^{-1}~pc^2}}\right) & = & 3.25\times10^7\left(\frac{\nu_\mathrm{rest}}{\mathrm{GHz}}\right)^{-2}(1+z)^{-1} \nonumber \\
& \times & \left(\frac{F}{\mathrm{Jy~km~s^{-1}}}\right)\left(\frac{d_L}{\mathrm{Mpc}}\right)^2,
\end{eqnarray}
where \(F\) is the total line flux, \(\nu_\mathrm{rest}\) is the rest frequency, \(z\) is the redshift, and \(d_L\) is the distance \citep{SDR92}. The luminosity is equivalent to the expression \(L'=A_0\sum_j W_j\), where \(A_0\) is the projected pixel area, \(W_j=\Delta v\sum_i T_{\mathrm{b},i}\) is the integrated intensity, \(T_\mathrm{b}\) is the brightness temperature, and summations are over \(j\)-th pixel and \(i\)-th velocity channel, where \(\Delta v\) is the channel width. Thus, \(L'\) is proportional to \(T_\mathrm{b}\) in this definition. To obtain luminosity in the units of \(L_\sun\), which is equivalent to the energy radiated away, one should use the equation \(L=1.04\times10^{-3}\nu_\mathrm{rest}(1+z)^{-1}Fd_L^2\) \citep{SDR92}. Inserting the values from Table \ref{tab:res} into this equation, we note that the luminosity \(L\) of [\ion{C}{1}] (1--0) is larger than that of CO (1--0) and CO (2--1); the line can contribute significantly to the cooling of interstellar gas.

\subsection{Molecular Gas Traced by CO (2--1)}\label{sec:cod}

\begin{figure*}
 \centering
  \includegraphics[width=1\textwidth]{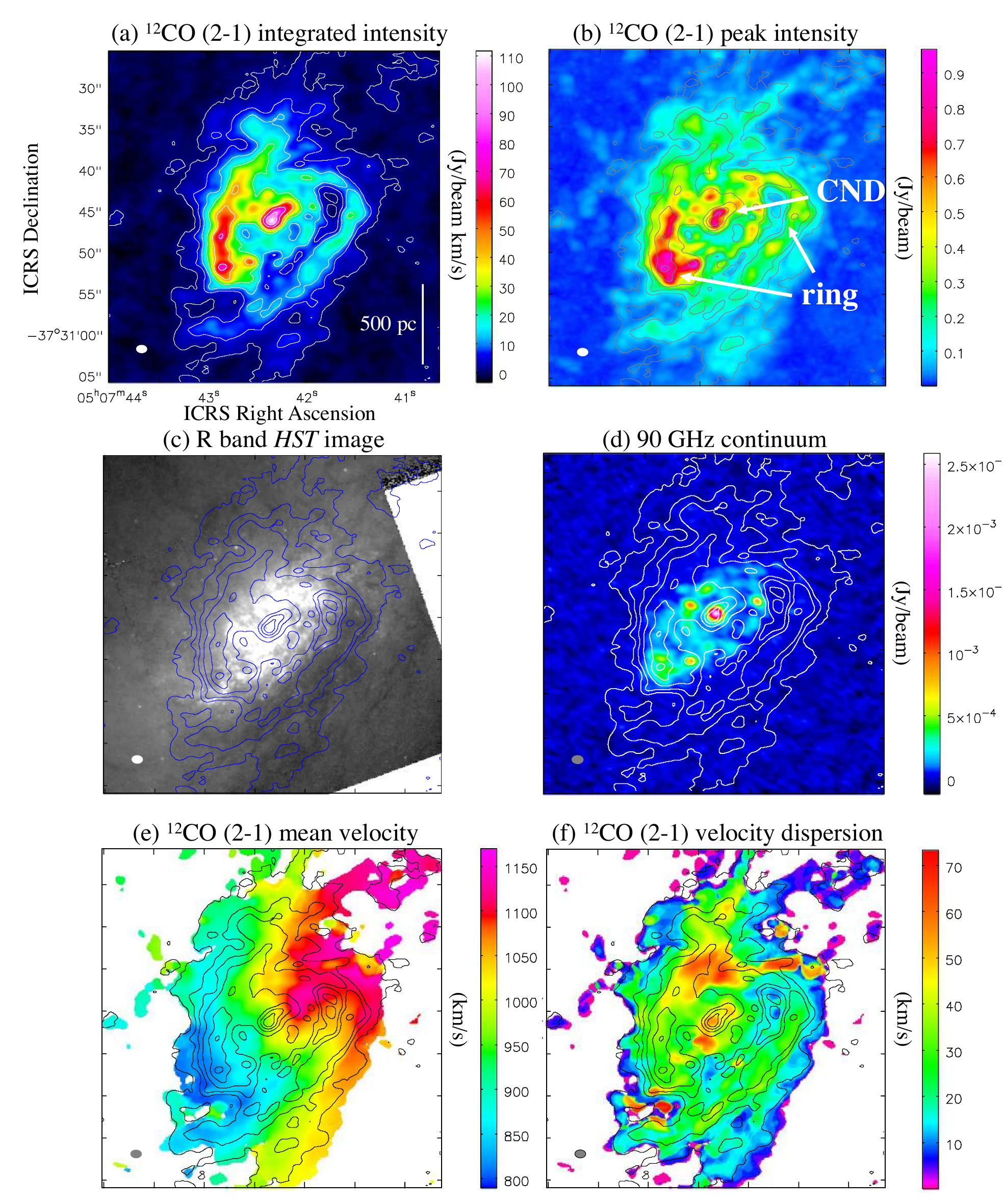}
 \caption{CO (2--1) images: (a) integrated intensity, (b) peak intensity, (c) 675W (R band) \emph{Hubble Space Telescope} image with integrated intensity contours (blue), (d) 658N (H\(\alpha\)) image with CO (2--1) contours, (e) mean velocity, and (f) velocity dispersion. The integrated intensity contours are plotted at \((0.01,0.05,0.1,0.2,0.4,0.6,0.8)\times112~\mathrm{Jy~beam^{-1}~km~s^{-1}}\) (maximum); \(1\sigma=0.50\) Jy beam\(^{-1}\) km s\(^{-1}\).
 The circumnuclear disk (CND) and the ring are marked in panel (b). The images in panels (e,f) are clipped below \(10\sigma\), where \(1\sigma=3.03~\mathrm{mJy~beam^{-1}}\). The beam size is shown at the bottom left corner. The southwest side of the disk is the near side.\label{fig:co}}
\end{figure*}

We begin by briefly describing the distribution and kinematics of molecular gas in NGC 1808 traced by the CO (2--1) emission. Due to its relatively low critical density (\(n_\mathrm{cr}\sim10^3~\mathrm{cm}^{-3}\)), the line is often used as a tracer of bulk molecular gas (e.g., \citealt{Ler13}). Figure \ref{fig:co}(a) shows the integrated intensity of the CO (2--1) line, defined as \(I\equiv \Delta v\sum_i S_i\), where \(S\) is the intensity in Jy beam\(^{-1}\) and the summation is over \(i\)-th velocity channel of width \(\Delta v\). ICRS stands for International Celestial Reference System, adopted in ALMA observations. The integrated intensity images were created in CASA using the task \emph{immoments} with no masking, yielding unbiased images. The maximum integrated intensity of \(I_\mathrm{max}=112.0\pm0.5\) Jy beam\(^{-1}\) km s\(^{-1}\) is found at \((\alpha,\delta)_\mathrm{ICRS}=(\mathrm{5^h7^m42\fs35,-37\arcdeg30\arcmin46\farcs3})\) (maximum pixel value) toward the galactic center in the region referred to as the circumnuclar disk (CND; marked in Figure \ref{fig:co}(b)), with an error of \(0\farcs9\) (beam size). The CND is a region abundant in dense molecular gas \citep{Sal18}; the star formation rate is of the order of \(\sim0.2~M_\sun~\mathrm{yr}^{-1}\) \citep{Bus17,Sal17}. The intensity is also prominent in the starburst ring, a structure composed of two major molecular spiral arms surrounding the CND at a radius of 400-500 pc, indicated in Figure \ref{fig:co}(b). The origin of the ring has been related to the location of the inner Lindblad resonance \citep{Sal16}. The molecular gas traced by CO (2--1) is also distributed throughout the disk between the CND and the ring in the form of giant molecular clouds (GMCs) and nuclear spiral arms, as well as beyond the ring at relatively low intensity (\(\lesssim10\%\) of the peak intensity). We refer to the central \(R<500\) pc region as the starburst disk.

In Figure \ref{fig:co}(b) we also show the peak intensity image (in Jy beam\(^{-1}\)), derived as the distribution of the maximum value of the spectrum in each image pixel, \(S_\mathrm{max}=\max[S(\alpha,\delta,v)]\), where \(\alpha\) and \(\delta\) are spatial coordinates, and \(v\) is LSR velocity. This image emphasizes high intensity in the starburst ring southeast of the galactic center, where we find a maximum value of \(S_\mathrm{max}=0.969\pm0.003\) Jy beam\(^{-1}\) at \((\alpha,\delta)_\mathrm{ICRS}=(\mathrm{5^h7^m42\fs57,-37\arcdeg30\arcmin51\farcs5})\), as well as weak emission in the outer regions where the line width is relatively narrow, hence faint when integrated over velocity.

A comparison of the CO (2--1) integrated intensity map with an optical image taken by the \emph{Hubble Space Telescope} of the central starburst region is shown in Figure \ref{fig:co}(c), where the R band traces the starlight. The starburst at optical wavelengths appears highly obscured by dust absorption. Figure \ref{fig:co}(d) shows the distribution of radio continuum at 90 GHz, a tracer of ionized gas that emits free-free radiation \citep{Sal17}. The distribution of the ionized gas continuum is notably different from CO in the central \(<500\) pc. We associate the continuum distribution with the starburst disk; it is the site of vigorous star formation activity and the ``hot spots'' reported in early studies (e.g., \citealt{Sai90}).

Figure \ref{fig:co}(e) shows the intensity-weighted mean velocity (moment 1) defined as \(\langle v\rangle\equiv\sum_i S_iv_i/{\sum_i S_i}\). This image was created by applying an intensity-based mask on the data cube to exclude pixels with low signal-to-noise ratio. The kinematics of CO gas is clearly dominated by rotation, with some evidence of large-scale noncircular motions, most notably in regions north and south of the center where the velocity field is S-shaped.

The velocity dispersion (moment 2), defined as \(\sigma_v=\sqrt{\sum_iS_i(v_i-\langle v\rangle)^2/{\sum_i S_i}}\), is shown in Figure \ref{fig:co}(f). The image was created by applying a mask on the data cube in the same way as for \(\langle v\rangle\) described above. This quantity can reveal regions of overlapping velocity components along the line of sight, e.g., due to extraplanar (outflow) gas motions. The overall distribution of \(\sigma_v\) is very similar to that previously observed in CO (1--0) and CO (3--2) data \citep{Sal17}. The typical velocity dispersion is of the order of \(\sim30~\mathrm{km~s}^{-1}\) throughout the central 1 kpc region at the resolution of 50 pc. The line widths (full width at half maximum, FWHM) in the GMCs around the CND and throughout the starburst ring are typically of the order of 20--40 km s\(^{-1}\), as estimated from fitting by a single Gaussian. In the CND, we measure large line widths of 70--100 km s\(^{-1}\) at the present resolution.

\subsection{[\ion{C}{1}] (1--0) and 500 GHz Continuum}\label{sec:cid}

\begin{figure*}
 \centering
  \includegraphics[width=1\textwidth]{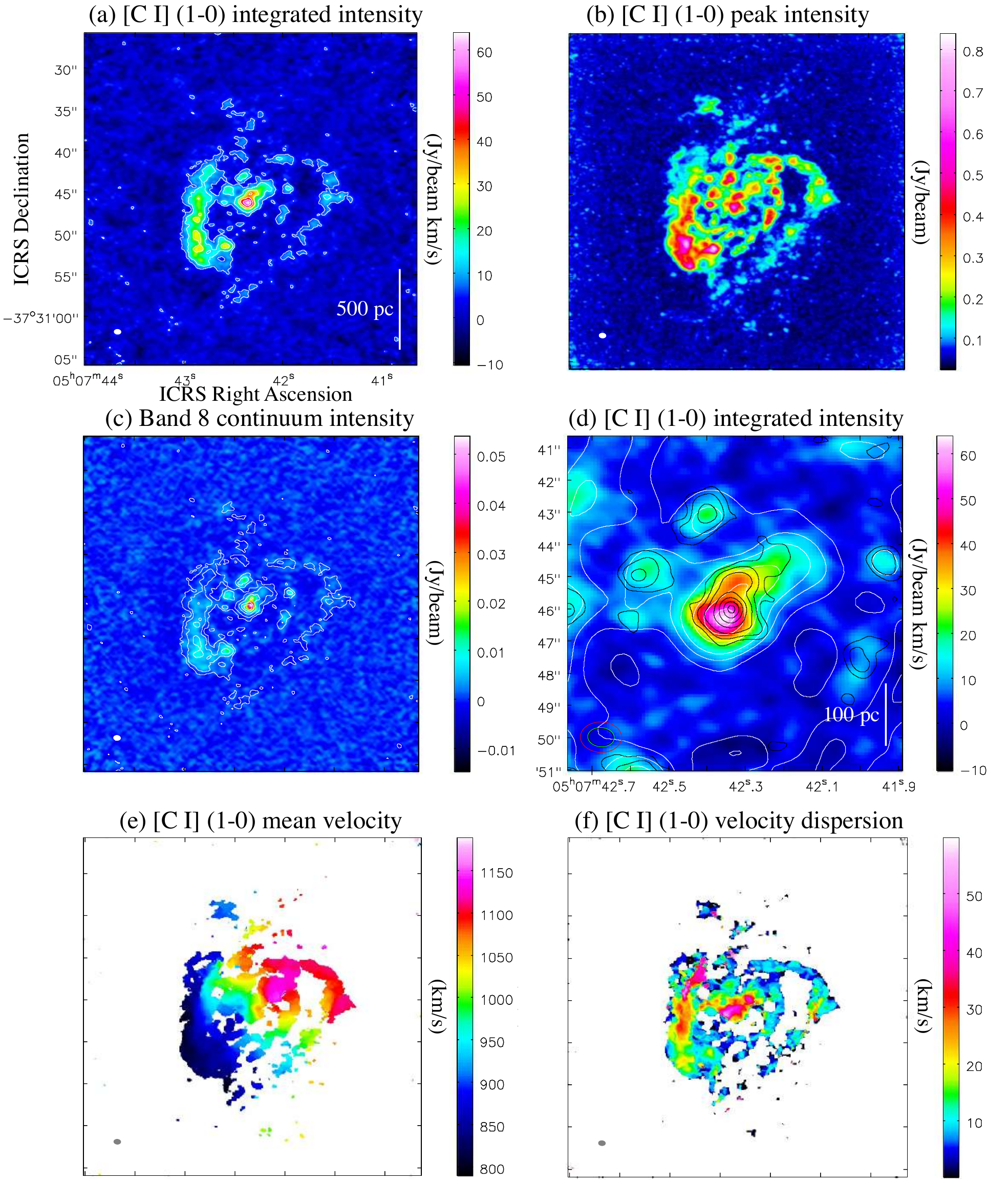}
 \caption{[\ion{C}{1}] (1--0) images: (a) integrated intensity, (b) peak intensity, (c) 500 GHz continuum with [\ion{C}{1}] integrated intensity contours, (d) [\ion{C}{1}] integrated intensity (color) in the CND with CO (2--1) contours (white) and 500 GHz continuum contours (black), (e) mean velocity, and (f) velocity dispersion. The contours in panels (a) and (c) are the [\ion{C}{1}] integrated intensity plotted at \((0.1,0.2,0.4,0.6,0.8)\times64.1~\mathrm{Jy~beam^{-1}~km~s^{-1}}\) (maximum); \(1\sigma=1.8\) Jy beam\(^{-1}\) km s\(^{-1}\). In panel (d), the CO (2--1) contours are \((0.01,0.05,0.1,0.2,0.3,0.4,0.6,0.8,0.95)\times112~\mathrm{Jy~beam^{-1}~km~s^{-1}}\) and the continuum contours are \((0.05,0.1,0.2,0.4,0.6,0.8,0.95)\times53.9~\mathrm{mJy~beam^{-1}}\); \(1\sigma=1.0\) mJy beam\(^{-1}\). The images in panels (e,f) are clipped below \(5\sigma\), where \(1\sigma=20.4~\mathrm{mJy~beam^{-1}}\). The beam size is shown at the bottom left corner.\label{fig:ci}}
\end{figure*}

Figure \ref{fig:ci} shows the distribution of [\ion{C}{1}] (1--0) in the central 1 kpc starburst region at 30 pc resolution. The neutral carbon emission is detected in all major structures, namely, in the CND and the ring as defined in Figure \ref{fig:co}. Although the spatial extent to which [\ion{C}{1}] (1--0) was detected is smaller than that of CO (2--1), the distributions appear to be similar. The maximum integrated intensity, shown in Figure \ref{fig:ci}(a), is \(I_\mathrm{max}=64\pm2\) Jy beam\(^{-1}\) km s\(^{-1}\) at \((\alpha,\delta)_\mathrm{ICRS}=(\mathrm{5^h7^m42\fs4,-37\arcdeg30\arcmin46\farcs3})\) (brightest pixel) inside the CND. The maximum intensity (Figure \ref{fig:ci}(b)) is \(S_\mathrm{max}=0.84\pm0.02\) Jy beam\(^{-1}\) at \((\alpha,\delta)_\mathrm{ICRS}=(\mathrm{5^h7^m42\fs56,-37\arcdeg30\arcmin51\farcs4})\) in the southeast part of the starburst ring.

The distribution of the 500 GHz continuum is shown in Figure \ref{fig:ci}(c). The maximum intensity is \(S_\mathrm{max}=54\pm1\) mJy beam\(^{-1}\) toward the CND. We estimated the coordinates of the continuum core by two-dimensional Gaussian fitting in a circular region of diameter \(1\arcsec\) centered at the brightest pixel. The resulting coordinates of the core are \((\alpha,\delta)_\mathrm{ICRS}=(\mathrm{5^h7^m42\fs34,-37\arcdeg30\arcmin46\farcs0})\) (error \(0\farcs6\)), consistent with recent high-resolution measurements at different frequencies, such as the continuum at 90 and 350 GHz \citep{Sal17,Com19}.

The neutral carbon emission closely follows that of CO (2--1) in the CND at the resolution of 30--50 pc (Figure \ref{fig:ci}(d)). The distribution of [\ion{C}{1}] (1--0) emission is also somewhat similar to that of the 500 GHz continuum in molecular clouds in the 1 kpc starburst disk, although [\ion{C}{1}] is relatively brighter in the arm east of the center. The location of the peak in the CND is different as well (Figure \ref{fig:ci}(d)); the internal structure of the CND is discussed in more detail in section \ref{spe}.

Figure \ref{fig:ci}(e,f) shows that the kinematics of [\ion{C}{1}] (1--0) resembles that of CO. The starburst disk is rotating and exhibits noncircular motions (S-shaped velocity field) in the central 300 pc, where \(\sigma_v\) is high.

\subsection{Dense gas tracers}\label{sec:dgt}

\subsubsection{\(^{13}\)CO (2--1)}\label{sec:cod13}

\begin{figure*}
 \centering
  \includegraphics[width=0.95\textwidth]{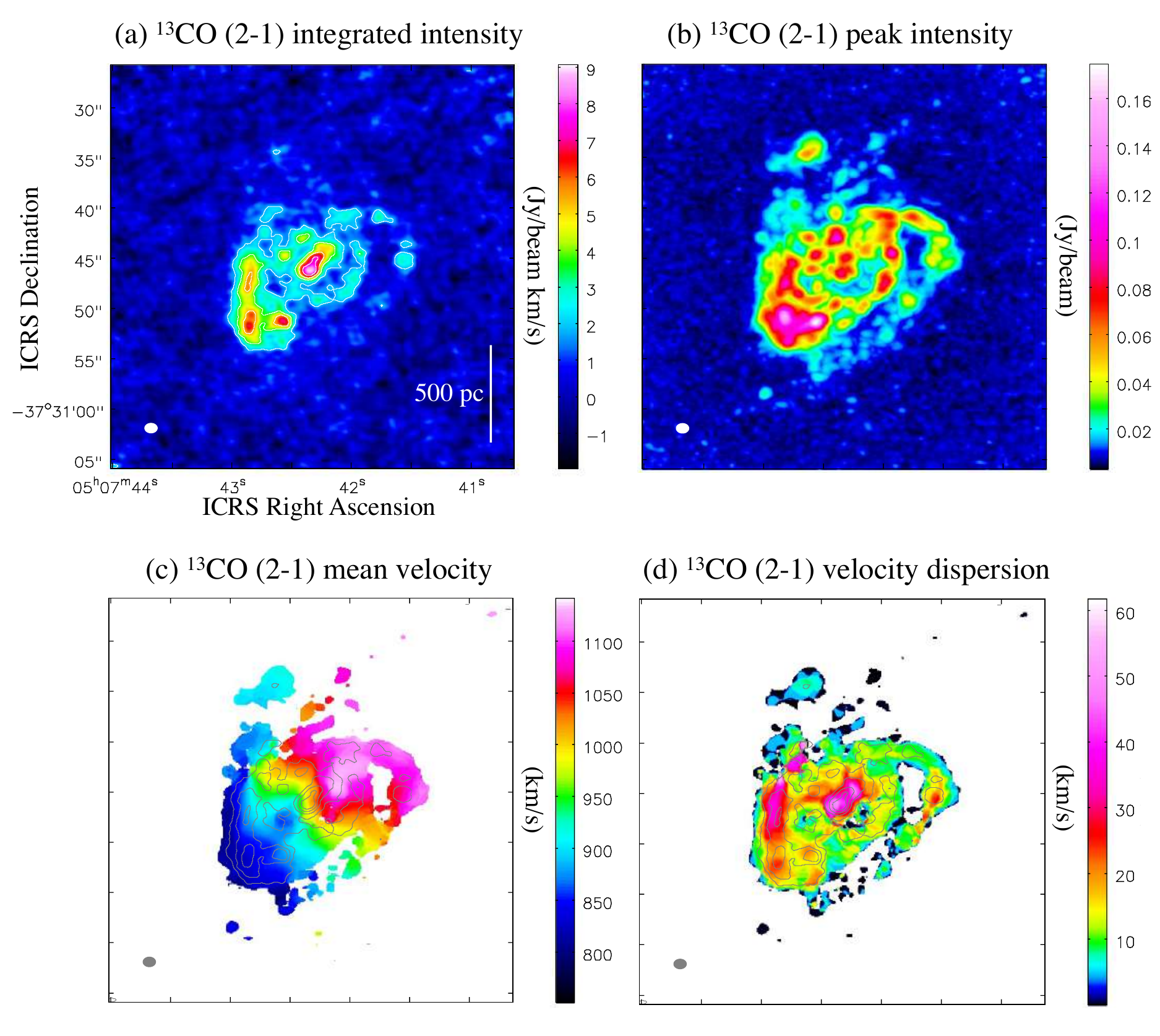}
 \caption{\(^{13}\)CO (2--1) images. The contours are the the integrated intensity at \((0.2,0.4,0.6,0.8)\times9.09~\mathrm{Jy~beam^{-1}~km~s^{-1}}\) (maximum); \(1\sigma=0.40\) Jy beam\(^{-1}\) km s\(^{-1}\). The images in panels (c) and (d) are clipped below \(5\sigma\), where \(1\sigma=2.84~\mathrm{mJy~beam^{-1}}\). The beam size is shown at the bottom left corner.\label{fig:den1}}
\end{figure*}

At a kinetic temperature of 20 K, \(^{13}\)CO (2--1) has a critical density of \(n_\mathrm{cr}/\beta\sim1\times10^4\) cm\(^{-3}\), and so traces moderately dense molecular gas. Figure \ref{fig:den1}(a-d) shows the integrated intensity, peak intensity, mean velocity, and velocity dispersion of \(^{13}\)CO (2--1), where the quantities are defined as before. Although \(^{13}\)CO is detected to a lower spatial extent, the overall structure, including the CND and starburst ring, is easily recognized. The most striking similarity appears to be with the [\ion{C}{1}] (1--0) intensity distribution (Figure \ref{fig:ci}) since both tracers were detected to a similar spatial extent.
The maximum integrated intensity is \(I_\mathrm{max}=9.1\pm0.5\) Jy beam\(^{-1}\) km s\(^{-1}\) at \((\alpha,\delta)_\mathrm{ICRS}=(\mathrm{5^h7^m42\fs35,-37\arcdeg30\arcmin46\farcs3})\) in the CND, which is the same as for CO (2--1). Figure \ref{fig:den1}(b) shows that the maximum intensity is not in the CND but at \((\alpha,\delta)_\mathrm{ICRS}=(\mathrm{5^h7^m42\fs57,-37\arcdeg30\arcmin51\farcs3})\) in the southeast part of the starburst ring, where we find \(S_\mathrm{max}=0.175\pm0.003\) Jy beam\(^{-1}\).

The gas kinematics traced by \(^{13}\)CO (2--1) is shown in Figure \ref{fig:den1}(c,d). Similar to CO (2--1) and [\ion{C}{1}] (1--0), the motion of \(^{13}\)CO gas is dominated by rotation in the galactic center region. The velocity dispersion is highest in the CND and eastern part of the starburst ring.

\subsubsection{\(C^{18}\)O (2--1)}

The C\(^{18}\)O (2--1) line is a tracer of relatively dense molecular gas, with a typical critical density of \(n_\mathrm{cr}/\beta\sim1\times10^4~\mathrm{cm}^{-3}\). The line tends to be optically thin, which makes it a good tracer of the interior structure of molecular clouds; however, due to relatively low protection against external UV radiation, the emission is typically confined to inner (denser) regions compared to those of CO and \(^{13}\)CO (2--1). Figure \ref{fig:den2}(a,b) indicates that C\(^{18}\)O (2--1) emission originates from a more compact region compared to \(^{13}\)CO (2--1). The fact that C\(^{18}\)O (2--1) is most strongly detected toward the CND suggests the presence of dense gas, and also that the excitation conditions are such that low-\(J\) levels are significantly populated, unlike in some active galactic nuclei (AGN), such as NGC 1068 and NGC 613, where C\(^{18}\)O (1--0) is not detected toward the CND, presumably due to high excitation \citep{Tak14,Miy18}. For example, high kinetic temperature in the center of NGC 1068, possibly related to shocks, has been reported by, e.g., \citet{GB10}, \citet{Kri11}, and \citet{Vit14}. The intensity maxima are \(I_\mathrm{max}=4.8\pm0.5\) Jy beam\(^{-1}\) km s\(^{-1}\) at \((\alpha,\delta)_\mathrm{ICRS}=(\mathrm{5^h7^m42\fs37,-37\arcdeg30\arcmin46\farcs3})\) in the CND, and \(S_\mathrm{max}=0.072\pm0.003\) Jy beam\(^{-1}\) at \((\alpha,\delta)_\mathrm{ICRS}=(\mathrm{5^h7^m42\fs58,-37\arcdeg30\arcmin51\farcs3})\) in the starburst ring.

\subsubsection{CS (5--4)}

Carbon monosulfide CS (5--4) emission was detected toward the CND (Figure \ref{fig:den2}(c)). With a critical density of \(n_\mathrm{cr}/\beta\sim5\times10^6~\mathrm{cm}^{-3}\) (at \(T_\mathrm{k}=20\) K), the line is a tracer of very dense molecular gas, that is expected to be limited to the compact interiors of clouds in the starburst nucleus. The CND is abundant in dense (\(n_\mathrm{H_2}\sim10^5~\mathrm{cm^{-3}}\)) molecular gas, making it the densest gas environment in NGC 1808 \citep{Sal18}. The measured intensity maxima are \(I_\mathrm{max}=4.4\pm0.4\) Jy beam\(^{-1}\) km s\(^{-1}\) at \((\alpha,\delta)_\mathrm{ICRS}=(\mathrm{5^h7^m42\fs33,-37\arcdeg30\arcmin46\farcs1})\), and \(S_\mathrm{max}=0.031\pm0.004\) Jy beam\(^{-1}\) at \((\alpha,\delta)_\mathrm{ICRS}=(\mathrm{5^h7^m42\fs35,-37\arcdeg30\arcmin45\farcs9})\). Unlike CO and [\ion{C}{1}], the peak intensity of CS (5--4) emission is in the CND and not in the starburst ring, though weak emission can be seen in the data cube toward the ring too.

\subsubsection{HNCO (10--9)}\label{sec:hnco}

\begin{figure*}
 \centering
  \includegraphics[width=0.95\textwidth]{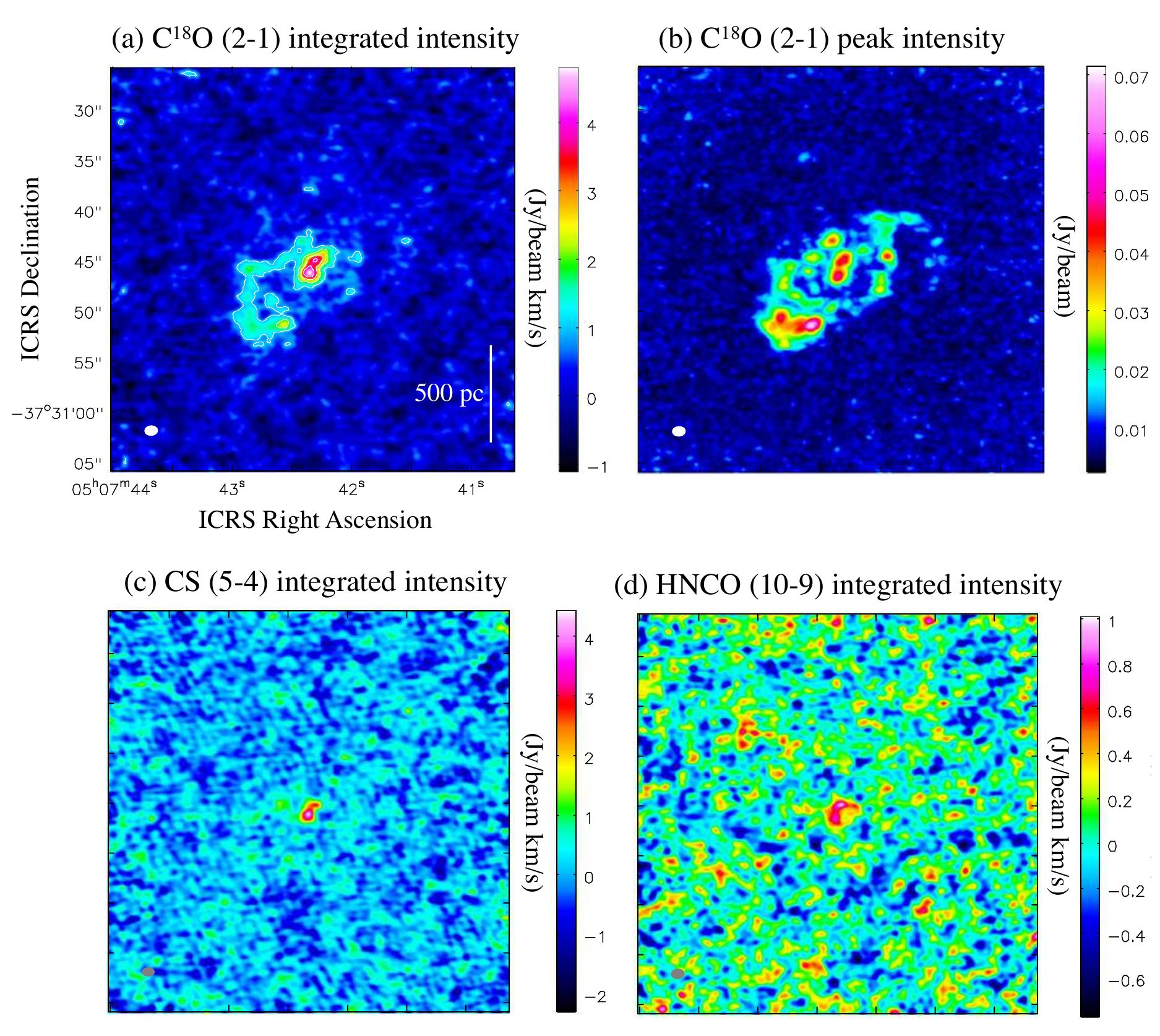}
 \caption{Dense gas tracers. The contours are plotted at \((0.2,0.4,0.6,0.8)\times4.82~\mathrm{Jy~beam^{-1}~km~s^{-1}}\) for C\(^{18}\)O; \(1\sigma=0.30\) Jy beam\(^{-1}\) km s\(^{-1}\).
 The beam size is shown at the bottom left corner.\label{fig:den2}}
\end{figure*}

The isocyanic acid HNCO (\(v=0,~J_{K_a,K_c}=10_{0,10}-9_{0,9}\)) was detected toward the CND (Figure \ref{fig:den2}(d)).
This is the second transition of HNCO detected in NGC 1808, following HNCO (\(v=0,~J_{Ka,Kc}=4_{0,4}-3_{0,3}\)) at 87.9 GHz reported in \citet{Sal18}, although the band 6 detection presented here is at a higher signal-to-noise ratio. A spectrum of HNCO (10-9) toward the CND is shown in Figure \ref{fig:spe}(f). The line is characteristic of relatively dense and/or warm gas, and may be a tracer of slow shocks, as indicated, e.g., in the studies of the protostar-associated outflow L1157 in the Galaxy and the AGN in NGC 1068 \citep{RF11,Kel17}. In the CND, the maximum integrated intensity is \(I_\mathrm{max}=1.1\pm0.2\) Jy beam\(^{-1}\) km s\(^{-1}\) and the maximum intensity in \(S_\mathrm{max}=0.019\pm0.003\) Jy beam\(^{-1}\). Unlike other tracers, the maxima of HNCO (10--9) emission are closer to the northern peak of the CND, at \((\alpha,\delta)_\mathrm{ICRS}=(\mathrm{5^h7^m42\fs30,-37\arcdeg30\arcmin44\farcs9})\) and \((\alpha,\delta)_\mathrm{ICRS}=(\mathrm{5^h7^m42\fs31,-37\arcdeg30\arcmin44\farcs9})\), respectively. The line was not detected in the starburst ring.

\section{Discussion}\label{sec:dis}

\subsection{Atomic Carbon in the CND}\label{sec:rat}

In this section, we analyze the internal structure of the CND using the high resolution images and line profiles presented in Figure \ref{fig:spe}.

The high-resolution CO (3--2) images presented in \citet{Aud17} and \citet{Com19} show that the continuum core, that harbors a gaseous torus of radius \(\sim6\) pc, lies in the middle of a two-arm spiral pattern the comprises the inner structure of the CND. In Figure \ref{fig:spe}(a), we present the [\ion{C}{1}] (1--0) integrated intensity image, created from the data taken by the 12-m array in order to achieve highest resolution (\(0\farcs648\times0\farcs473\)). Also plotted are CO (3--2) at resolution \(0\farcs314\times0\farcs302\) as white contours and 500 GHz continuum as dashed black contours. These CO (3--2) data were acquired from the ALMA Archive (project 2016.1.00296.S) and used only here for comparison because of high angular resolution; the CO (3--2) data presented elsewhere in this paper are our cycle 2 data corrected for missing flux \citep{Sal17}. The distributions show that [\ion{C}{1}] (1--0) emission is similarly present in the nuclear spiral arms, following the CO (3--2) distribution at a scale of \(\sim25\) pc. Note that the high-resolution CO (3--2) image and our 500 GHz image exhibit a peak at the location of the core (AGN torus; see \citealt{Com19}), whereas the peak of [\ion{C}{1}] at lower resolution is \(\sim0\farcs5\) southeast of the core. A higher-resolution [\ion{C}{1}] image is needed to compare the two morphologies on small scales in the core.

In Figure \ref{fig:spe}(b), we show the [\ion{C}{1}] gas kinematics in the central 500 pc (mean velocity \(\langle v\rangle\)). Note that the kinematic position angle (direction perpendicular to the isovelocity curves) in the CND region (within the dashed circle) is \(\sim270\arcdeg\), whereas the position angle of the starburst disk and the bulge is \(310\arcdeg\mathrm{-}324\arcdeg\) \citep{Sal16}. The P.A. of the AGN torus (radius \(6\pm2\) pc) is \(245\arcdeg\pm8\arcdeg\) \citep{Com19}, which is comparable to the P.A. of the entire CND and different from the galactic disk. This distortion, which is also seen in CO (3-2) images \citep{Sal17,Com19} and near-infrared images of ionized gas and hot H\(_2\) gas \citep{Bus17}, may be a result of inflow motions at radii \(<100\) pc, or a warped nuclear disk with respect to the galactic disk.

\subsubsection{Line Intensities and Profiles}\label{spe}

\begin{figure*}
 \centering
  \includegraphics[width=0.8\textwidth]{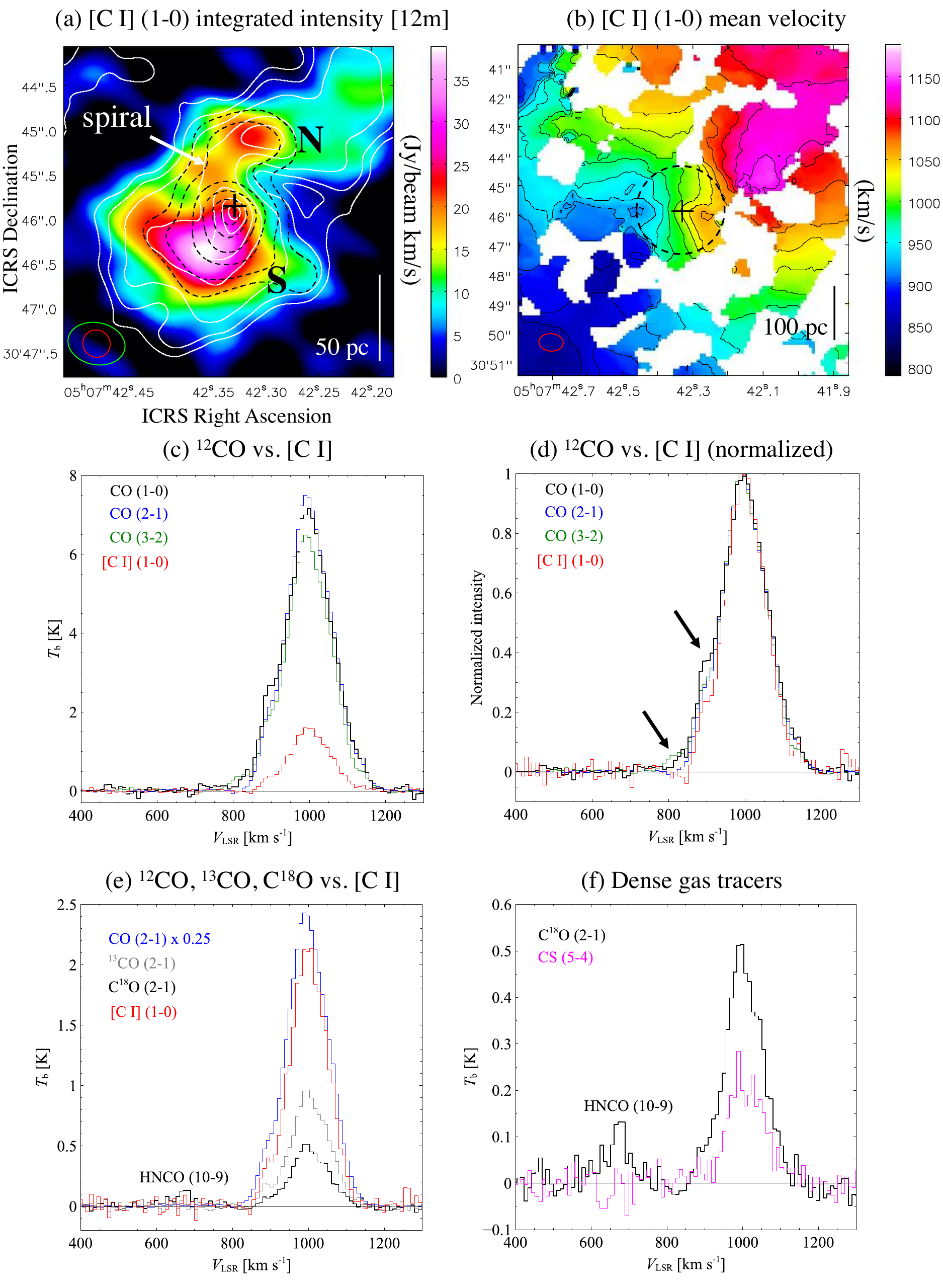}
 \caption{(a) Circumnuclear disk (CND). The north and south peaks are denoted by ``N'' and ``S'', and the spiral arm pattern is indicated by an arrow. The plus symbol is the center (core) from Table \ref{tab:gal}. The white contours are CO (3--2) at \((0.05,0.1,0.2,0.4,0.6,0.8,0.95)\times I_\mathrm{max}\) at resolution \(0\farcs314\times0\farcs302\) (from ALMA Archive); the dashed black contours are 500 GHz continuum (12m) at \((0.1,0.2,0.4,0.6,0.8,0.95)\times45.4\) mJy beam\(^{-1}\); \(1\sigma=1.0\) mJy beam\(^{-1}\). (b) Mean velocity field (as in Figure \ref{fig:ci}) with isovelocity contours from 860 to 1040 km s\(^{-1}\) in steps of 20 km s\(^{-1}\). (c-f) Mean spectra toward the CND (diameter \(3\arcsec\)). The data were smoothed to the angular resolution of CO (1--0) in panels (c,d). In panels (e,f), they are in their original format. The arrows in panel (d) indicate a blueshift bump and wing (see text). The image in panel (a) is from the 12-m array data only, whereas the image in panel (b) and the spectra were made from total flux images. The CO (3--2) spectra in panels (c,d) were produced from our cycle 2 data.}\label{fig:spe}
\end{figure*}

In order to analyze the physical conditions, we derive the basic properties of the mean spectra toward an aperture of \(3\arcsec\) diameter that encloses the CND (Figure \ref{fig:spe}, Table \ref{tab:cnd}). The figure shows a comparison between CO (1--0), CO (2--1), CO (3--2), \(^{13}\)CO (2--1), C\(^{18}\)O (2--1), CS (5--4), and [\ion{C}{1}] (1--0), where the intensity scale \(T_\mathrm{b}\) is expressed as the Rayleigh-Jeans brightness temperature. The CO (1--0) and CO (3--2) data used here are from \citet{Sal17}.

\begin{deluxetable*}{lcccccc}
\tablecaption{Line Parameters of Mean CND Spectra\label{tab:cnd}}
\tablecolumns{7}
\tablewidth{0pt}
\tablehead{
\colhead{Parameter} &
\colhead{CO (1--0)} &
\colhead{CO (2--1)} &
\colhead{CO (3--2)} &
\colhead{\(^{13}\)CO (2--1)} &
\colhead{C\(^{18}\)O (2--1)} &
\colhead{[\ion{C}{1}] (1--0)}
}
\startdata
\(T_\mathrm{b}\) (K) & \(7.119\pm0.079\) & \(7.449\pm0.024\) & \(6.441\pm0.027\) & \(0.710\pm0.019\) & \(0.398\pm0.021\) & \(1.594\pm0.042\) \\
\(V_\mathrm{c}\) (km s\(^{-1}\)) & \(996.85\pm0.77\) & \(998.32\pm0.43\) & \(996.33\pm0.64\) & \(1003.95\pm0.98\) & \(1003.1\pm1.9\) & \(1000.74\pm0.78\) \\
\(V_\mathrm{FWHM}\) (km s\(^{-1}\)) & \(155.8\pm1.8\) & \(145.7\pm1.0\) & \(146.1\pm1.5\) & \(139.1\pm2.3\) & \(126.6\pm4.5\) & \(130.6\pm1.8\) \\
\(W\) (K km s\(^{-1}\)) & \(1109.9\pm6.7\) & \(1108.2\pm2.1\) & \(975.4\pm2.1\) & \(97.2\pm1.6\) & \(47.4\pm1.4\) & \(214.7\pm3.5\) \\
\enddata
\tablecomments{\(T_\mathrm{b}\) is the maximum value. The center velocity \(V_\mathrm{c}\) and the line width \(V_\mathrm{FWHM}\) were calculated by simple Gaussian fitting using CASA's \emph{Spectral Profile} tool; the effective uncertainties are set by the channel width \(\Delta v=9.5\) km s\(^{-1}\) and \(V_\mathrm{c}\) for all six lines are within this error. The uncertainty of \(W\) is \(\Delta T_\mathrm{b}\sqrt{\Delta v \Delta v_\mathrm{b}}\), where \(\Delta v_\mathrm{b}\) is the baseline width (emission-free channels) and \(\Delta T_\mathrm{b}\) is the rms noise over the baseline. Only statistical uncertainties are stated; flux calibration uncertainty (10-15\%) is not included. The mean spectra were obtained within a circular aperture of diameter \(3\arcsec\) toward the CND (angular resolution \(2\farcs666\times1\farcs480\)).}
\end{deluxetable*}

\begin{figure*}
 \centering
  \includegraphics[width=0.5\textwidth]{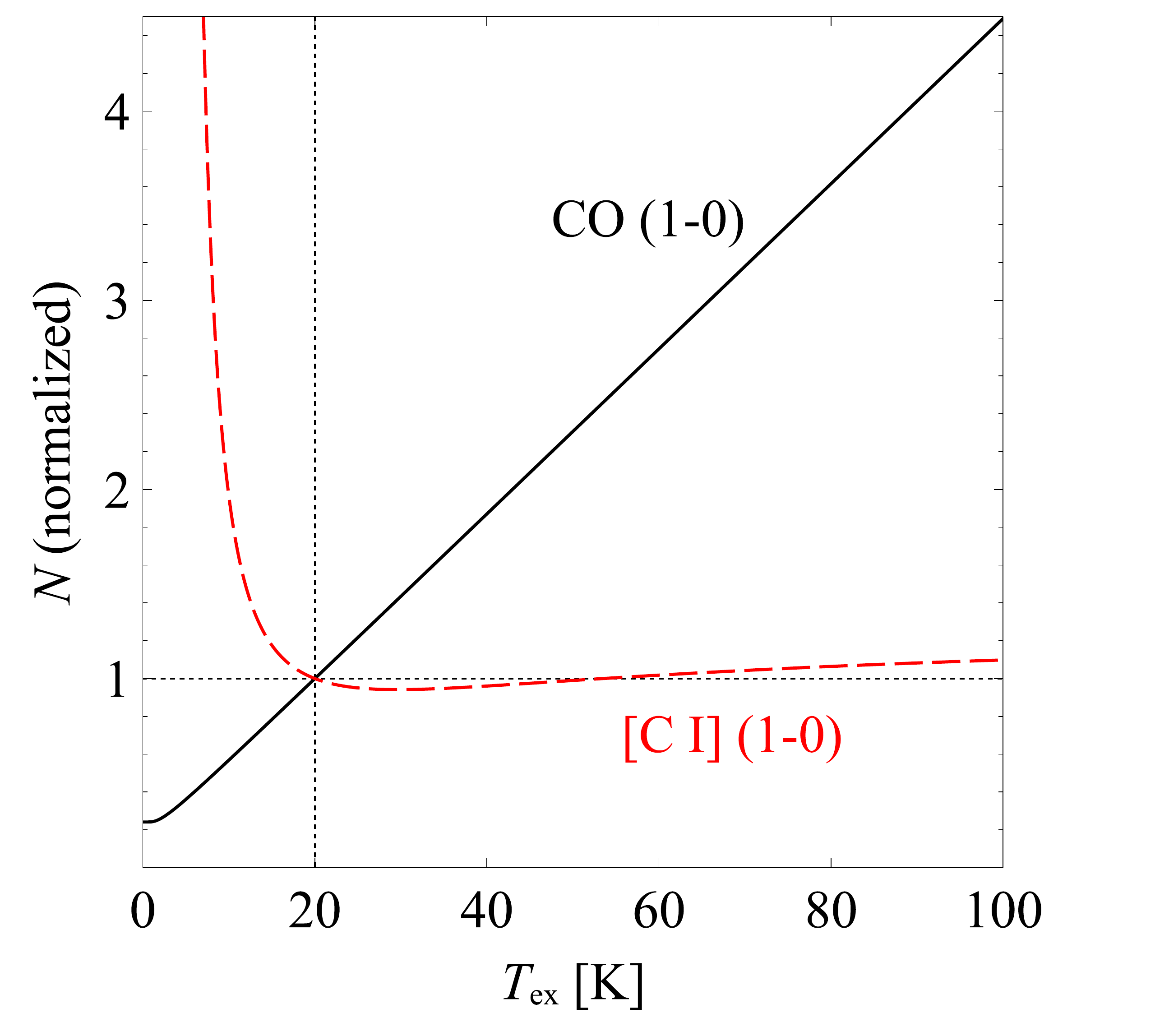}
 \caption{Column density \(N\) from equations \ref{comass} and \ref{cimass} as a function of excitation temperature (normalized to unity for \(T_\mathrm{ex}=20\) K) for CO (1--0) (solid black curve) and [\ion{C}{1}] (1--0) (dashed red).\label{fig:plt}}
\end{figure*}

Figure \ref{fig:spe}(c) shows the [\ion{C}{1}] (1--0) and low-\(J\) \(^{12}\)CO lines, smoothed to the common resolution of \(\sim100\) pc, corresponding to the CO (1--0) data. The smoothed images were derived by convolving the original ones with Gaussian kernels using the CASA task \emph{imsmooth}. We find that the three \(^{12}\)CO lines have very similar mean intensities of \(\sim7\) K (within flux calibration error of 10\%), whereas the intensity of [\ion{C}{1}] (1--0) line is \(\sim22\%\) that of CO (1--0) toward the CND. In panel (d), the spectra are normalized so that we can compare the profile shapes. The lines are generally similar, except for the blueshift bump (at \(V_\mathrm{LSR}\sim900~\mathrm{km~s^{-1}}\)) and wing (at \(V_\mathrm{LSR}\sim800~\mathrm{km~s^{-1}}\)) marked with arrows where the CO intensity is relatively larger than that of [\ion{C}{1}] (1--0). At the resolution of 100 pc, any differences that may exist in the distributions of [\ion{C}{1}] and CO in the continuum core appear to be smoothed out.

Figure \ref{fig:spe}(e) and Table \ref{tab:cnd} also reveal that the line profiles and widths of \(^{13}\)CO (2--1) and C\(^{18}\)O (2--1) in the CND are generally similar to that of [\ion{C}{1}] (1--0) at the resolution of \(\sim50\) pc, albeit little narrower than those of the low-\(J\) lines of \(^{12}\)CO. This may be caused partially by the relative difference of \(^{12}\)CO and [\ion{C}{1}] intensities in the bump and wing, since the FWHM line widths were determined by fitting a single Gaussian. The derived peak intensities and integrated intensities include only statistical uncertainties. This result suggests that these tracers are present in all major structures in the CND. For comparison, the similarity between CO, \(^{13}\)CO (2--1), and [\ion{C}{1}] (1--0) line profiles has also been reported for star forming regions in the Large Magellanic Cloud \citep{Oka19}. We also find that [\ion{C}{1}] (1--0) is brighter than \(^{13}\)CO (2--1), in agreement with single-dish measurements of a number of galactic nuclei in \citet{IB02}.

\subsubsection{Column Density and Abundance}\label{abu}

The CO (1--0) integrated intensity toward the CND can give us an estimate of the column density and total mass of H\(_2\) gas. We begin by calculating the column density of CO. The lower limit is given by simplified conditions of local thermodynamic equilibrium (LTE), where the kinetic temperature is equal to the excitation temperature (\(T_\mathrm{k}=T_\mathrm{ex,CO}\)), and optically thin (optical depth \(\tau\ll1\)) CO (1--0) emission as

\begin{eqnarray}\label{comass}
N_\mathrm{CO}
& = & N_1\frac{Q}{g_1}\mathrm{e}^{E_1/{kT_\mathrm{ex,CO}}} \nonumber \\
& = & \frac{3k}{4\pi^3\mu^2\nu_\mathrm{CO}}\left(1-\mathrm{e}^{-E_1/kT_\mathrm{ex,CO}}\right)^{-1}W_\mathrm{CO},
\end{eqnarray}
where \(N_1\) is the column density of the \(J=1\) rotational level, \(E_1/k=5.53\) K is the energy of the level, \(Q=\sum_{J=0}^\infty g_J\mathrm{e}^{-hBJ(J+1)/{kT_\mathrm{ex,CO}}}=kT_\mathrm{ex,CO}/hB\) is the partition function (where \(B\) is the rotational constant), \(g_J\) is the statistical weight, \(\nu_\mathrm{CO}\) is the frequency of the transition, \(\mu=0.112\) D is the dipole moment, and \(W_\mathrm{CO}\equiv\Delta v\sum T_\mathrm{b}\) is the measured CO (1--0) integrated intensity (Table \ref{tab:cnd}). Assuming an CO/H\(_2\) abundance ratio of \(10^{-4}\), we get the H\(_2\) gas column density as \(N_\mathrm{H_2}\sim(1\mathrm{-}4)\times10^{22}~\mathrm{cm}^{-2}\) for \(T_\mathrm{ex,CO}=20\mathrm{-}100\) K. The dependence on \(T_\mathrm{ex}\) is shown in Figure \ref{fig:plt}.

On the other hand, using a Galactic CO-to-H\(_2\) conversion factor of \(X_\mathrm{CO}=2\times10^{20}~\mathrm{cm^{-2}(K~km~s^{-1})^{-1}}\) \citep{Bol13}, we obtain \(N_\mathrm{H_2}=X_\mathrm{CO}W_\mathrm{CO}\sim2\times10^{23}~\mathrm{cm}^{-2}\). Using a low conversion factor of \(X_\mathrm{CO}=0.5\times10^{20}~\mathrm{cm^{-2}(K~km~s^{-1})^{-1}}\), which may be more appropriate for starburst galaxies \citep{Bol13}, the column density is \(N_\mathrm{H_2}\sim5\times10^{22}~\mathrm{cm}^{-2}\). The total H\(_2\) gas mass is given by \(M_\mathrm{H_2}=2m_\mathrm{p}N_\mathrm{H_2}\mathcal{A}\), where \(m_\mathrm{p}\) is the proton mass, and \(\mathcal{A}\) is the projected area of the sampled region, and we find \(M_\mathrm{H_2}\sim1.5\times10^7~M_\sun\) in the CND region (diameter \(3\arcsec\)). The total molecular gas mass is \(M_\mathrm{mol}=1.36M_\mathrm{H_2}\sim2\times10^7M_\sun\), where the factor 1.36 is the correction for the abundance of helium and other elements.

Similarly, assuming LTE and optically thin [\ion{C}{1}] (1--0) emission, the column density of atomic carbon gas can be derived (see Appendix \ref{app:A}) from

\begin{equation}\label{cimass}
N_\mathrm{CI}
=\frac{8\pi k\nu_\mathrm{[CI]}^2}{hc^3A_\mathrm{[CI]}}\frac{Q}{g_1}\mathrm{e}^{E_1/{kT_\mathrm{ex,[CI]}}}W_\mathrm{[CI]},
\end{equation}
where \(\nu_\mathrm{[CI]}\) is the frequency of the transition, \(Q=1+3\mathrm{e}^{-E_1/{kT_\mathrm{ex,[CI]}}}+5\mathrm{e}^{-E_2/{kT_\mathrm{ex,[CI]}}}\) is the partition function, \(g_J\) is the statistical weight, \(A_\mathrm{[CI]}=7.93\times10^{-8}\) s\(^{-1}\) is the Einstein coefficient\footnote{From NIST Atomic Spectra Database Lines Data: \url{https://physics.nist.gov/asd}.} and \(E_1/k=23.6\) K and \(E_2/k=62.5\) K are the energies of the \({^3}\mathrm{P}_1\) and \({^3}\mathrm{P}_2\) levels, respectively. The total mass of atomic carbon gas is \(M_\mathrm{CI}=m_\mathrm{C}N_\mathrm{CI}\mathcal{A}\), where \(m_\mathrm{C}\) is the carbon atom mass.

For a range of \(T_\mathrm{ex,[CI]}=20\mathrm{-}100\) K, we use the value of \(W_\mathrm{[CI]}\) in Table \ref{tab:cnd} and obtain \(N_\mathrm{CI}\sim3\times10^{18}\) cm\(^{-2}\). The dependence on \(T_\mathrm{ex,[CI]}\) is weak in this range (Figure \ref{fig:plt}) and we get \(M_\mathrm{CI}\sim5\times10^3~M_\sun\) for all values between 20 and 100 K. For \(T_\mathrm{ex,[CI]}=50\) K, the mass is \(M_\mathrm{CI}=(5.4\pm0.1)\times10^3~M_\sun\), where the error includes only the statistical uncertainty of the integrated intensity \(W_\mathrm{[CI]}\). Thus, the mass ratio becomes \(M_\mathrm{CI}/M_\mathrm{H_2}\sim3.5\times10^{-4}\) and the mean \ion{C}{1}/H\(_2\) abundance ratio in the CND is \(N_\mathrm{CI}/N_\mathrm{H_2}=M_\mathrm{CI}/(6M_\mathrm{H_2})\sim6\times10^{-5}\). Toward a two-pixel aperture of maximum [\ion{C}{1}] (1--0) and CO (1--0) integrated intensities at \(\sim2\arcsec\) resolution, we find \(W_\mathrm{[CI]}/W_\mathrm{CO}=0.22\), and the abundance ratio becomes \(\sim7\times10^{-5}\). These values are consistent within a factor of two with the estimates of \(\sim3\times10^{-5}\) for the Cloverleaf quasar at redshift 2.5 \citep{Wei03}, \((3.9\pm0.4)\times10^{-5}\) in submillimeter galaxies (SMGs) \citep{AZ13}, \((8.4\pm3.5)\times10^{-5}\) in SMGs and quasars at redshift 2.5 \citep{Wal11}, \(7\times10^{-5}\) in dusty star-forming galaxies at redshift 4 \citep{Bot17}, \(2\times10^{-5}\) in the main-sequence galaxies at redshift 1.2 \citep{Val18}, and \((2.5\pm1.0)\times10^{-5}\) as the average in a sample of nearby galaxies that includes starbursts and AGN \citep{Jia19}. These authors used similar methods to calculate the mass. The result is also similar to \(N_\mathrm{CI}\sim1\mathrm{-}3\times10^{18}\) cm\(^{-2}\) found in the central disk of the starburst galaxy NGC 253 \citep{Kri16}, who applied the derivation method from \citet{Ike02}. Note also that the \ion{C}{1}/CO abundance is \(\sim0.5\) in the CND, which is \(\sim5\) times larger than in the Orion cloud \citep{Ike02}, and a factor of 2 larger than the abundance in the bulk gas of the Galactic Central Molecular Zone, albeit comparable to some extreme clouds there \citep{Tan11}.

\subsubsection{Non-LTE Calculations of Physical Conditions}\label{sec:rad}

Are LTE and optically thin line reasonable assumptions for [\ion{C}{1}] (1--0) toward the CND? To verify this condition we ran a series of calculations using the non-LTE radiative transfer program RADEX \citep{vdT07}. The velocity width was fixed at \(\Delta V=25\) km s\(^{-1}\), which is reasonable since such extreme clouds have been observed in the Galactic center (e.g., \citealt{Oka01a}), and the background temperature to 2.73 K. We varied the column density \(N\) (over three orders of magnitude), kinetic temperature \(T_\mathrm{k}\), and density \(n_\mathrm{H_2}\). Since \(\tau\propto N/\Delta V\) in the equations used by RADEX, the calculations are sensitive only to this ratio. Varying \(N\) affects the resulting physical conditions (typical variations are a factor of \(\sim2\) in temperature and dex \(<1\) in density), and the ratio \(N/\Delta V\) that yields a solution where all investigated line intensity ratios intersect in a narrow range of the temperature-density parameter space is regarded as closest to the actual conditions. The geometry was set to be an expanding sphere, with an escape probability of \(\beta=(1-\mathrm{e}^{-\tau})/\tau\), equivalent to the large velocity gradient (LVG) approximation \citep{Sob57,GK74,SS74}.

\begin{figure*}
 \centering
  \includegraphics[width=0.75\textwidth]{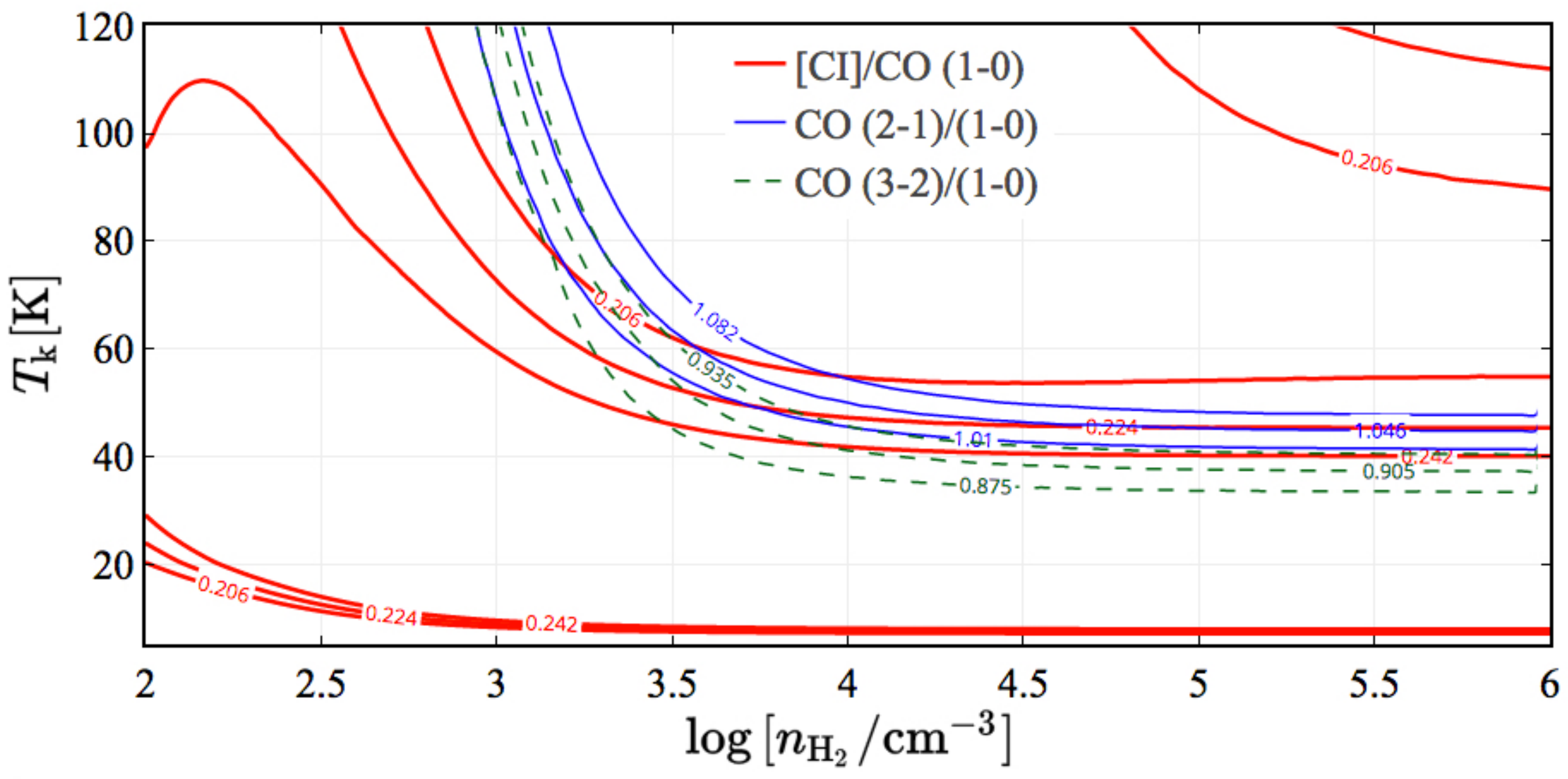}
 \caption{An example of LVG calculations for the CND where it was assumed that [\ion{C}{1}] and CO (1--0) emissions arise from the same regions. Here, \(N_\mathrm{CO}=6.0\times10^{18}\) cm\(^{-2}\), \(N_\mathrm{CI}=3.5\times10^{18}\) cm\(^{-2}\), and \(\Delta V=25\) km s\(^{-1}\). The middle curves are the observed ratios; the outer curves are \(\pm3\sigma\) (not including flux calibration uncertainty).\label{fig:lvg}}
\end{figure*}

\begin{deluxetable*}{lccc}
\tablecaption{LVG Calculations of [\ion{C}{1}] and CO Line Excitation Temperatures and Opacities\label{tab:rad}}
\tablecolumns{6}
\tablewidth{0pt}
\tablehead{
\colhead{\(N\) (cm\(^{-2}\))} &
\colhead{\(3\times10^{18}\)} &
\colhead{\(3\times10^{17}\)} &
\colhead{\(3\times10^{16}\)}
}
\startdata
\(T_\mathrm{k}=50\) K, \(n_\mathrm{H_2}=10^2\) cm\(^{-3}\) \\
\(T_\mathrm{ex}\) (K) & \textbf{14} [16, 15, 11] & \textbf{11} [7.1, 5.9, 5.1] & \textbf{10} [4.2, 3.9, 5.6] \\
\(\tau\) & \textbf{1.2} [10, 23, 24] & \textbf{0.15} [3.4, 4.8, 1.2] & \textbf{0.02} [0.67, 0.45, 0.04] \\ \hline
\(T_\mathrm{k}=50\) K, \(n_\mathrm{H_2}=10^3\) cm\(^{-3}\) \\
\(T_\mathrm{ex}\) (K) & \textbf{34} [32, 30, 26] & \textbf{31} [23, 14, 11] & \textbf{31} [20, 7.9, 8.1] \\
\(\tau\) & \textbf{0.40} [2.7, 9.0, 14] & \textbf{0.04} [0.64, 2.5, 2.5] & \textbf{0.004} [0.09, 0.44, 0.19] \\ \hline
\(T_\mathrm{k}=50\) K, \(n_\mathrm{H_2}=10^4\) cm\(^{-3}\) \\
\(T_\mathrm{ex}\) (K) & \textbf{49} [49, 45, 44] & \textbf{49} [99, 35, 28] & \textbf{49} [460, 31, 20] \\
\(\tau\) & \textbf{0.22} [1.2, 4.4, 7.5] & \textbf{0.02} [0.08, 0.75, 1.3] & \textbf{0.002} [0.002, 0.10, 0.19] \\ \hline
\(T_\mathrm{k}=20\) K, \(n_\mathrm{H_2}=10^2\) cm\(^{-3}\) \\
\(T_\mathrm{ex}\) (K) & \textbf{10} [11, 10, 7.3] & \textbf{7.7} [5.6, 4.6, 4.0] & \textbf{7.5} [3.6, 3.4, 4.5] \\
\(\tau\) & \textbf{1.7} [17, 34, 24] & \textbf{0.20} [4.7, 5.0, 0.76] & \textbf{0.02} [0.80, 0.41, 0.03] \\ \hline
\(T_\mathrm{k}=20\) K, \(n_\mathrm{H_2}=10^3\) cm\(^{-3}\) \\
\(T_\mathrm{ex}\) (K) & \textbf{16} [17, 17, 15] & \textbf{15} [12, 9.6, 7.1] & \textbf{15} [8.8, 5.8, 6.1] \\
\(\tau\) & \textbf{1.0} [7.7, 19, 21] & \textbf{0.11} [1.6, 3.6, 2.3] & \textbf{0.01} [0.26, 0.51, 0.13] \\ \hline
\(T_\mathrm{k}=20\) K, \(n_\mathrm{H_2}=10^4\) cm\(^{-3}\) \\
\(T_\mathrm{ex}\) (K) & \textbf{20} [20, 20, 19] & \textbf{19} [20, 17, 15] & \textbf{19} [24, 14, 11] \\
\(\tau\) & \textbf{0.80} [5.8, 15, 17] & \textbf{0.08} [0.66, 2.0, 2.2] & \textbf{0.01} [0.06, 0.26, 0.24] \\
\enddata
\tablecomments{The values in boldface are for [\ion{C}{1}] (1--0), and those in [ ] are for CO (1--0), (2--1), (3--2), respectively. The line width is \(\Delta V=25\) km s\(^{-1}\).}
\end{deluxetable*}

Examples of the radiative transfer calculations are shown in Figure \ref{fig:lvg} and Table \ref{tab:rad}. From the investigated parameter space, we find that the conditions that are close to the assumptions and LTE results above are those of relatively warm (\(T_\mathrm{k}\sim40\mathrm{-}80\) K) and moderately dense (\(n_\mathrm{H_2}\sim10^{3\mathrm{-}4}\) cm\(^{-3}\)) gas. The observed brightness temperature ratios of CO (3--2)/CO (1--0), CO (2--1)/CO (1--0), and [\ion{C}{1}] (1--0)/CO (1--0) as in Figure \ref{fig:spe} are reproduced within \(3\sigma\) when the column densities are set to \(N_\mathrm{CI}=3.5\times10^{18}\) cm\(^{-2}\) and \(N_\mathrm{CO}=6.0\times10^{18}~\mathrm{cm^{-2}}\) for a common velocity width of \(\Delta V=25\) km s\(^{-1}\). This column density \(N_\mathrm{CI}\) is equivalent to the LTE result when \(\tau_\mathrm{[CI]}\approx0.3\), which yields a correction factor of \(\tau_\mathrm{[CI]}/(1-\mathrm{e}^{-\tau_\mathrm{[CI]}})\approx1.16\) to the optically thin value (see equation \ref{cdci} in Appendix \ref{app:A}); this is in agreement with \(\tau_\mathrm{[CI]}\) derived by RADEX. Strictly speaking, the comparison of brightness temperatures between [\ion{C}{1}] (1--0) and CO (1--0) makes sense only under the assumption that the emissions originate from the same region inside molecular clouds with the same physical conditions of H\(_2\) gas.

The optical depth of CO (1--0) under these conditions is \(\tau\sim2\) and the lines are nearly thermalized (\(T_\mathrm{ex,CO(1-0)}\approx T_\mathrm{ex,[CI]}\approx T_\mathrm{k}\)). Other low-\(J\) lines are also nearly thermalized and have higher optical depths. The \(^{13}\)CO (2--1) line is optically thin and subthermally excited. These results suggest that the abundance of \ion{C}{1} is enhanced and comparable (\(\sim0.5\)) to that of CO in the CND.

The estimated physical conditions are similar to those in the center of M82. \citet{Stu97} constrained the temperature and density of the emitting gas to \(T_\mathrm{k}\gtrsim50\) K and \(n_\mathrm{H_2}\sim10^4\) cm\(^{-3}\) from the line intensity ratio of [\ion{C}{1}] (2--1)/(1--0), measured by a single dish telescope. The beam-averaged column density \(N_\mathrm{CI}\) toward the center of M82 was reported to be \(\sim2\times10^{18}\) cm\(^{-2}\) \citep{Sch93,Stu97}. Similarly, in Orion A, as a representative star-forming Galactic cloud, \citet{Shi13} found \(\tau_\mathrm{[CI]}<1\), comparable \(T_\mathrm{ex}\), and column densities in the range \(\sim10^{17\mathrm{-}18}\) cm\(^{-2}\), with highest values toward regions such as Orion KL. 

The results of calculations also give us an insight into the behavior of [\ion{C}{1}] (1--0) optical depth in various physical conditions. Compared to the CND, the gas density and temperature in the more quiescent galaxy disk are expected to be generally lower, with \(n_\mathrm{H_2}\sim10^{2\mathrm{-}3}\) cm\(^{-3}\) and \(T_\mathrm{k}\lesssim20\) K. Table \ref{tab:rad} tells us that the [\ion{C}{1}] line is in some cases marginally optically thick and subthermally excited. For example, taking \(N_\mathrm{CI}=3\times10^{16-17}\) cm\(^{-2}\) and \(\Delta V=2.5\) km s\(^{-1}\) results in \(\tau_\mathrm{[CI]}=0.2\mathrm{-}2\), where the upper limit corresponds to a high \(N_\mathrm{CI}\) and low \(T_\mathrm{k}\). When the density is high (\(n_\mathrm{H_2}=10^4\) cm\(^{-3}\)), the [\ion{C}{1}] line is thermalized.

\subsection{Atomic Carbon in the Central 1 kpc Starburst Disk}\label{tbr}

\subsubsection{Intensity Ratios}\label{sec:radpro}

\begin{figure*}
 \centering
  \includegraphics[width=0.85\textwidth]{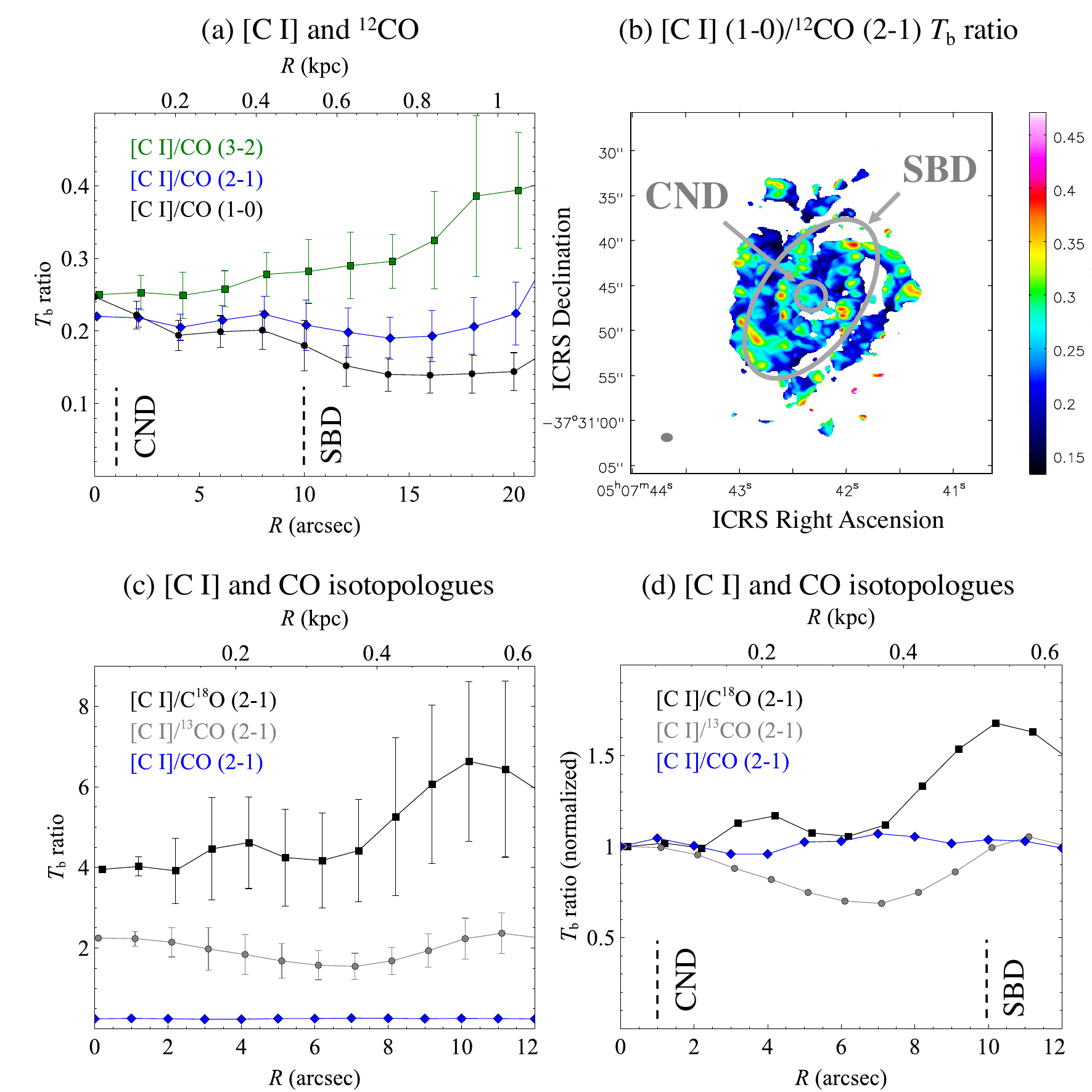}
 \caption{Azimuthally-averaged radial profiles of \(T_\mathrm{b}\) ratios derived for position angle \(PA=324\arcdeg\) and inclination \(i=57\arcdeg\). The data were sampled in elliptical rings in radial steps of \(2\arcsec\) in the top panels (low resolution) and \(1\arcsec\) in the bottom panels (high resolution). The points of different data sets are separated by \(0\farcs1\). The error bars are r.m.s. (not shown in panel (d) for clarity). The vertical dashed lines indicate the radii of the CND (1\arcsec) and starburst disk (SBD; radius 10\arcsec). The ratio image in panel (b) is clipped below \(2\sigma\) of the [\ion{C}{1}] image, where \(1\sigma=0.30~\mathrm{K}\).
 \label{fig:tbr}}
\end{figure*}

In this section, we investigate the spatial variation of the [\ion{C}{1}]/CO line intensity ratio in the central 1 kpc starburst region, as an indicator of excitation conditions. The ratios are presented on \(T_\mathrm{b}\) [K] scale in Figure \ref{fig:tbr} as azimuthally-averaged radial profiles at a resolution of \(\sim100\) pc in panel (a) and \(\sim50\) pc in panels (c,d). The position angle and inclination adopted for the geometry of elliptical rings are from Table \ref{tab:gal}. We also show the [\ion{C}{1}] (1--0)/CO (2--1) intensity ratio in panel (b); the ratio maps for all tracers are given in Figure \ref{fig:tbr1} in Appendix \ref{app:B}; the images show that the intensity ratios are approximately axisymmetric, i.e., there are no large changes with respect to azimuthal angle. The major trends are described below.

(1) The [\ion{C}{1}] (1--0)/CO (1--0) intensity ratio (denoted by \(r_{10}\)) is high (\(\sim0.20\mathrm{-}0.25\)) in the CND and \(R\lesssim500\) pc, which is the starburst disk defined by strong 93 GHz continuum emission in Figure \ref{fig:co}(d) and discussed in section \ref{sec:res}. The ratio decreases outward to reach \(r_{10}\sim0.15\) at larger radii, comparable to the typical ratios in the Galactic disk. A similar non-uniform ratio is observed also toward the starburst galaxy NGC 253 \citep{Kri16}. The mean intensity ratio over the rings in the central radius \(20\arcsec\) region is \(0.18\pm0.04\). The error, \(1\sigma\) of the mean, is a measure of variation in the radial direction.

(2) The [\ion{C}{1}] (1--0)/CO (2--1) intensity ratio (\(\equiv r_{21}\)) is notably uniform throughout the central 1 kpc: the mean value over the rings within a radius of \(20\arcsec\) is \(r_{21}=0.21\pm0.01\). There is no major difference between the CND, starburst disk within \(R\lesssim500\) pc, and outer regions, as is clear from Figure \ref{fig:tbr}(b), which shows the spatial distribution of \(r_{21}\) at a resolution of \(\sim50\) pc. Although there are local variations typically \(\Delta r_{21}\sim0.05\), there are no global gradients. This is possibly a result of two effects. First, the excitation energies of the upper levels of the two lines (16.6 K for CO and 23.6 K for [\ion{C}{1}]; Table \ref{tab:res}) are relatively similar. Second, while [\ion{C}{1}] (1--0) is often optically thin (\(\tau\lesssim1\); Table \ref{tab:rad}), CO (2--1) is almost certainly optically thick (\(\tau>1\)) in most regions. In that case, the critical densities of the two lines can become very similar.

To illustrate this, we consider the critical density at \(T_\mathrm{k}=20\) K: \(n_\mathrm{cr,CO(2-1)}/\beta_\mathrm{CO(2-1)}\sim1\times10^4\) cm\(^{-3}\) for CO (2--1) and \(n_\mathrm{cr,[CI]}/\beta_\mathrm{[CI]}\sim1\times10^3\) cm\(^{-3}\) for [\ion{C}{1}] (1--0). Then, if \(\tau_\mathrm{CO(2-1)}=10\), expanding sphere (LVG) geometry yields \(\beta_\mathrm{CO(2-1)}=0.1\) and \(n_\mathrm{cr,CO(2-1)}\approx n_\mathrm{cr,[CI]}\). This effect of radiative trapping can bring the two critical densities to comparable values.

Interestingly, \citet{Val18} found that the \(L'_\mathrm{[CI]}/L'_\mathrm{CO(2-1)}\) ratio is approximately constant among unresolved high redshift objects that include main-sequence and starburst galaxies (although with large scatter). Despite the enormous difference in spatial scales (resolved central 1 kpc vs. unresolved entire galaxies), the two results are consistent with each other and indicate the \(r_{21}\) may be least sensitive to galactic environment.

\begin{figure}
 \centering
  \includegraphics[width=0.5\textwidth]{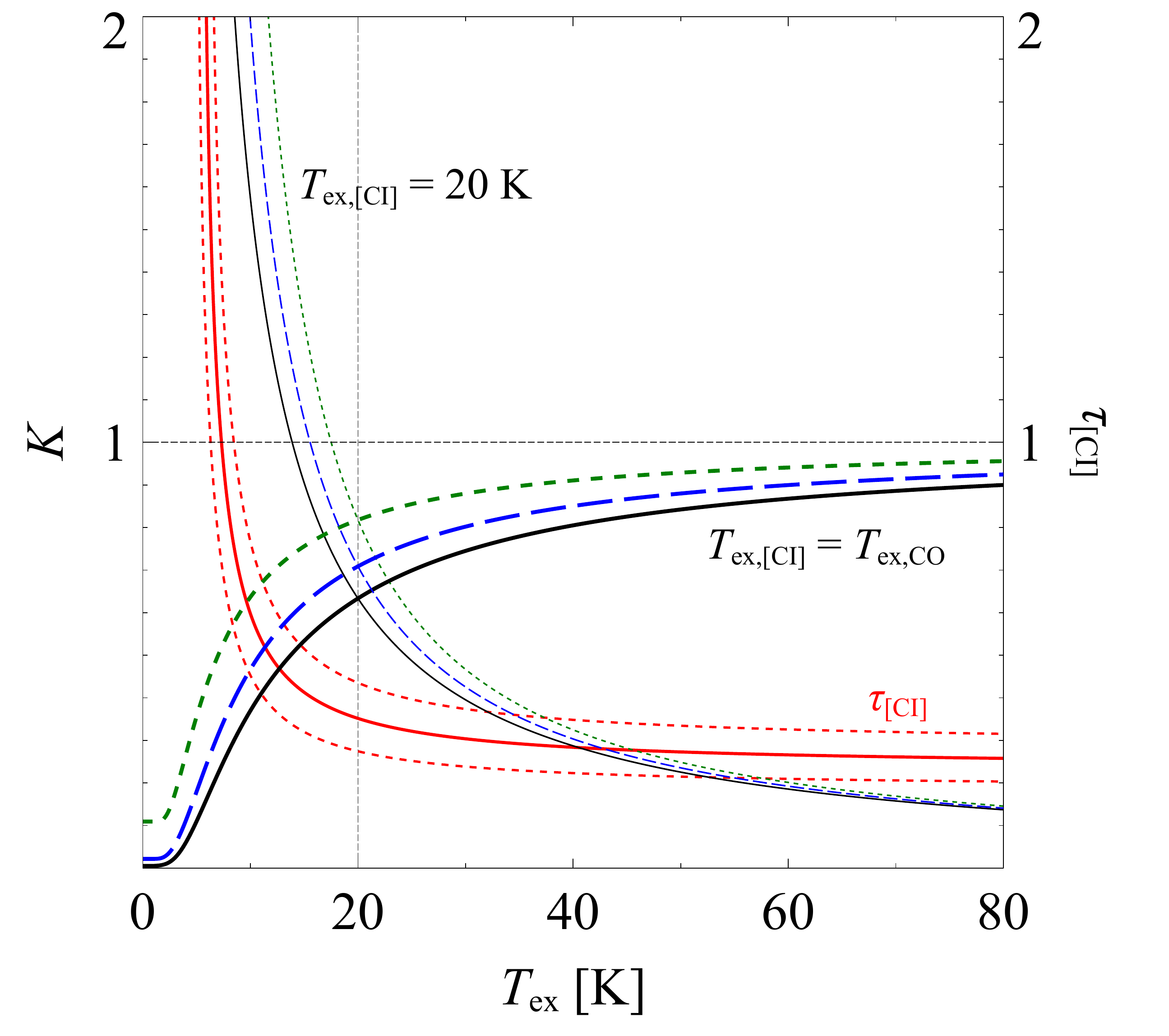}
 \caption{Thick curves: \(K\) factor from equation \ref{rte} when \(T_\mathrm{ex,CO}=T_\mathrm{ex,[CI]}\) for CO (1--0) (solid black curve), CO (2--1) (dashed blue), and CO (3--2) (dotted green). The resulting optical depth of [\ion{C}{1}] (1--0) is plotted for \(r_{21}=0.21\pm0.04\) as red curves. Thin curves: \(K\) factor when \(T_\mathrm{ex,[CI]}=20\) K and \(T_\mathrm{ex,CO}\) is a free parameter.}\label{fig:plt0}
\end{figure}

(3) The [\ion{C}{1}] (1--0)/CO (3--2) intensity ratio (\(\equiv r_{32}\)) is uniform in the central 300 pc, gradually increases at larger radii, and the maximum values are observed toward the edge of the starburst disk. The mean ratio over the rings within a radius of \(20\arcsec\) is \(0.30\pm0.05\).

(4) The intensity ratios of [\ion{C}{1}] (1--0) with those of \(^{13}\)CO (\(\equiv r^{13}_{21}=1.92\pm0.26\) within a radius of \(10\arcsec\)) and C\(^{18}\)O (2--1) (\(\equiv r^{18}_{21}=4.70\pm0.86\)) are plotted in Figure \ref{fig:tbr}(c,d), where the errors are \(1\sigma\). The error bars of \(r^{18}_{21}\) are largest partially because the correlation with [\ion{C}{1}] is poor (section \ref{sec:llc}), and increase at large radii because the intensity of C\(^{18}\)O is relatively weak. For comparison, the relative errors in Figure \ref{fig:tbr}(c) are \(16\%\), \(21\%\), and \(26\%\) for the curves of \(r_{21}\), \(r^{13}_{21}\), and \(r^{18}_{21}\), respectively, averaged within the radius of \(10\arcsec\). Note that, unlike \(r_{21}\), which is uniform across the region, \(r^{13}_{21}\) exhibits a decline by \(\sim30\%\) between the CND (where \(\sim2.2\)) and the ring. The ratio is higher than in the Galactic clouds, and \(\sim2\)  times lower than the average in nearby starbursts found by \citet{IRv15} for local starbursts using low-resolution data. On the other hand, \(r^{18}_{21}\) is mostly uniform at radii \(R\lesssim0.4\) kpc, and increases sharply beyond the starburst disk, where C\(^{18}\)O (2--1) emission is weak. This trend is expected if the average gas density decreases in the outer regions.

\subsubsection{Excitation and Optical Depth}

The origin of the observed line intensity ratios can be investigated using radiative transfer equations. The ratio of the measured (background subtracted) [\ion{C}{1}] (1--0) and CO intensities is

\begin{eqnarray}\label{rte}
r&\equiv& \frac{J(T_\mathrm{ex,[CI]})-J(T_\mathrm{cmb})}{J(T_\mathrm{ex,CO})-J(T_\mathrm{cmb})}\frac{1-\mathrm{e}^{-\tau_\mathrm{[CI]}}}{1-\mathrm{e}^{-\tau_\mathrm{CO}}} \notag \\
& = & K \frac{1-\mathrm{e}^{-\tau_\mathrm{[CI]}}}{1-\mathrm{e}^{-\tau_\mathrm{CO}}},
\end{eqnarray}
where \(J(T_\mathrm{ex})=(h\nu/k)(\mathrm{e}^{h\nu/kT_\mathrm{ex}}-1)^{-1}\) is the radiation temperature and ``cmb'' is the cosmic microwave background. First, let us assume that \(T_\mathrm{ex,CO(2-1)}\approx T_\mathrm{ex,[CI]}\) (see Table \ref{tab:rad}). Since CO (2--1) is optically thick in most conditions, \(\tau_\mathrm{[CI]}\) can be derived as

\begin{equation}\label{tau}
\tau_\mathrm{[CI]}\approx-\ln\left(1-\frac{r_{21}}{K}\right).
\end{equation}
For the observed ratio of \(r_{21}=0.21\), the optical depth is \(\tau_\mathrm{[CI]}\approx0.35\) at \(T_\mathrm{ex}=20\) K and only weakly depends on \(T_\mathrm{ex}\) (Figure \ref{fig:plt0}). Note that we do not derive \(\tau_\mathrm{[CI]}\) directly from \(T_\mathrm{b,[CI]}\) using the method in \citet{Oka01b} and \citet{Ike02}, because the beam filling factor is unknown and may be \(<1\), in which case the observed \(T_\mathrm{b}\) underestimates the actual \(\tau\); our derivation only assumes that the beam filling factors of CO (2--1) and [\ion{C}{1}] (1--0) are equal. From equation \ref{tau}, we deduce that the observed uniform intensity ratio can be a consequence of a relatively uniform, low (\(\lesssim1\)) opacity of [\ion{C}{1}] (1--0). Test calculations using RADEX suggest that the assumption of nearly equal excitation temperatures of CO (2--1) and [\ion{C}{1}] (1--0) holds for typical conditions that pervade in the central 1 kpc: \(N_\mathrm{CO}/\Delta V=1\times10^{17}\) cm\(^{-2}\) (km s\(^{-1}\))\(^{-1}\), \(N_\mathrm{CI}=(0.1\mathrm{-}1)N_\mathrm{CO}\), \(T_\mathrm{k}=20\) K, and \(n_\mathrm{H_2}=10^{3\mathrm{-}4}\) cm\(^{-3}\). RADEX also yields \(\tau_\mathrm{CO(2-1)}\gg1\) for most conditions listed in Table \ref{tab:rad}.

By comparison, even though CO (1--0) is easily thermalized and often \(T_\mathrm{ex,CO(1-0)}\approx T_\mathrm{ex,[CI]}\) (Table \ref{tab:rad}), the \(r_{10}\) ratio exhibits a gradient. This is expected if the optical depth of CO (1--0) is relatively low (\(\sim1\)) in the inner regions so that the rightmost term in equation \ref{rte} increases. On the other hand, \(r_{32}\) is approximately constant in the central 300 pc, similar to \(r_{21}\), and then increases outward. The increase could be a consequence of a combination of excitation and optical depth effects. The opacity of CO (3--2) is large, but its excitation temperature decreases below thermalization level at low densities (Table \ref{tab:rad}). Subthermal excitation contributes to an increase of \(r_{32}\) in the outer regions; the trend is illustrated by thin curves in Figure \ref{fig:plt0}, where we note a steep increase of \(K\) when \(T_\mathrm{ex,CO}<T_\mathrm{ex,[CI]}\).

In general, the \(^{13}\)CO (2--1) and C\(^{18}\)O (2--1) lines are often subthermally excited, and the excitation temperatures are different from that of [\ion{C}{1}]. Assuming that \(T_\mathrm{ex,^{13}\mathrm{CO
}}=T_\mathrm{ex,\mathrm{C^{18}O}}\) and that both lines are optically thin, the intensity ratio becomes \(T_{\mathrm{b},^{13}\mathrm{CO
}}/T_{\mathrm{b,C^{18}O}}\approx\tau_{^{13}\mathrm{CO
}}/\tau_{\mathrm{C^{18}O}}\approx N_{^{13}\mathrm{CO
}}/N_{\mathrm{C^{18}O}}\). The observed azimuthally-averaged ratio is 2--3 in the starburst disk.


\subsubsection{Abundance and Mass in the Central 1 kpc}

\begin{figure}
 \centering
  \includegraphics[width=0.45\textwidth]{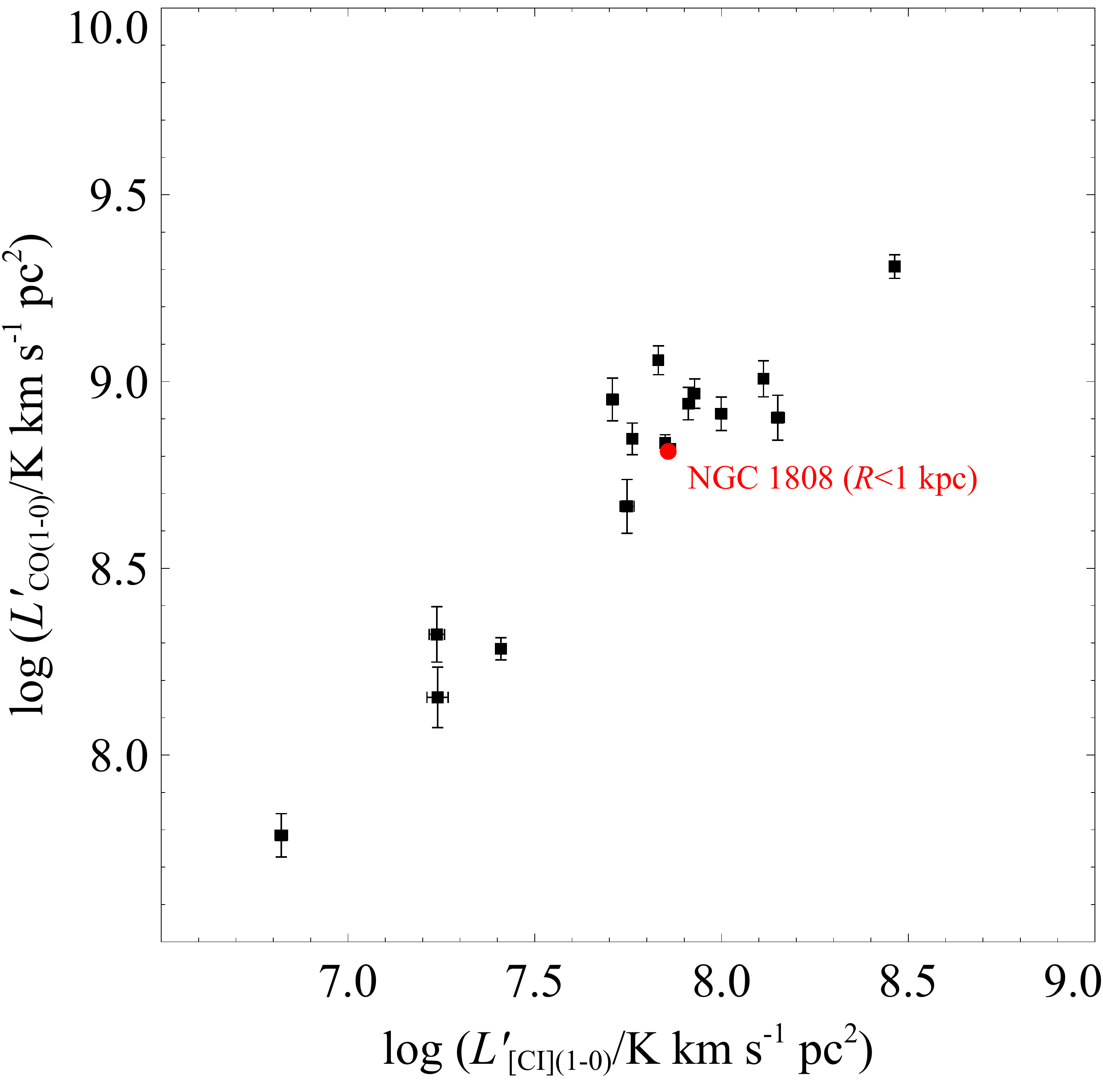}
 \caption{[\ion{C}{1}] (1--0) and CO (1--0) total luminosities of 15 nearby galaxies from \cite{Jia19}. The red circle is NGC 1808 from this work.\label{fig:lum}}
\end{figure}

Assuming optically thin emission, we use equation \ref{cimass} and luminosity from Table \ref{tab:res} and calculate the total \ion{C}{1} mass in the central 1 kpc as \(M_\mathrm{CI}\sim9.7\times10^4~M_\sun\) for a range of \(T_\mathrm{ex,[CI]}=20\mathrm{-}50\) K. The total H\(_2\) gas mass is calculated as \(M_\mathrm{H_2}=\alpha_\mathrm{CO}L'_\mathrm{CO(1-0)}/1.36\sim5.2\times10^8~M_\sun\), where the conversion factor \(\alpha_\mathrm{CO}=0.25\times4.3~M_\sun~(\mathrm{K~km~s^{-1}~pc^{-2}})^{-1}\) is the recommended value for the Galactic center and starbursts \citep{Bol13}. The factor 1/1.36 is multiplied to subtract the contribution from helium and heavy elements. The \ion{C}{1}/H\(_2\) abundance is then \(\sim3\times10^{-5}\), a factor of two lower than in the CND and consistent with the values in nearby and distant galaxies (see section \ref{abu}). The luminosity ratio \(L'_\mathrm{[CI]}/L'_\mathrm{CO(1-0)}=0.11\pm0.02\) is also in agreement with the median value of \(0.11\pm0.04\) for 15 nearby galaxies at 1 kpc resolution \citep{Jia19}. A comparison of the CO and [\ion{C}{1}] luminosities in the central 1 kpc region of NGC 1808 with the total luminosities of nearby galaxies is shown in Figure \ref{fig:lum}.

\subsubsection{\(L'_\mathrm{CO}\mathrm{-}L'_\mathrm{[CI]}\) Correlations}\label{sec:llc}

The potential of [\ion{C}{1}] (1--0) as a tracer of total molecular gas mass has been investigated recently by measuring its correlation with CO (1--0). \citet{Jia19} have shown that, for a sample of 15 nearby galaxies, including Seyfert and starburst galaxies, at \(\sim1\) kpc resolution, the intensities of the two lines can be expressed using a near-linear relation of \(\log{L'_\mathrm{CO(1-0)}}=(0.74\pm0.12)+(1.04\pm0.02)\log{L'_\mathrm{[CI](1-0)}}\), where \(L'\) is as defined by equation \ref{eq:lum}. By adding ultra-luminous infrared galaxies (ULIRGs) from \citet{Jia17} and high-redshift objects from \citet{Emo18} to the sample of nearby galaxies, the relation retains its near-linear nature. The total luminosities in the central 1 kpc in NGC 1808 also agree with this result (Figure \ref{fig:lum}). However, due to lack of angular resolution in previous studies, the behavior of the relation in different environments within individual galaxies has not yet been clarified. Moreover, \citet{IRv15} analyzed a sample of starbursts and (U)LIRGs and concluded that [\ion{C}{1}] may be tracing predominantly dense (\(10^4\) cm\(^{-3}\)) gas.

We now investigate the effect of environment on the \(L'_\mathrm{CO}\)--\(L'_\mathrm{[CI](1-0)}\) relations. To enable direct comparison, the data are presented as integrated intensity \(W\) [K km s\(^{-1}\)], which is proportional to luminosity as the quantity used to estimate H\(_2\) mass from CO observations. The integrated intensity ratio images of [\ion{C}{1}] (1--0) and all five CO lines are given in Figure \ref{fig:wr} in Appendix \ref{app:B}.

\begin{figure*}
 \centering
  \includegraphics[width=0.95\textwidth]{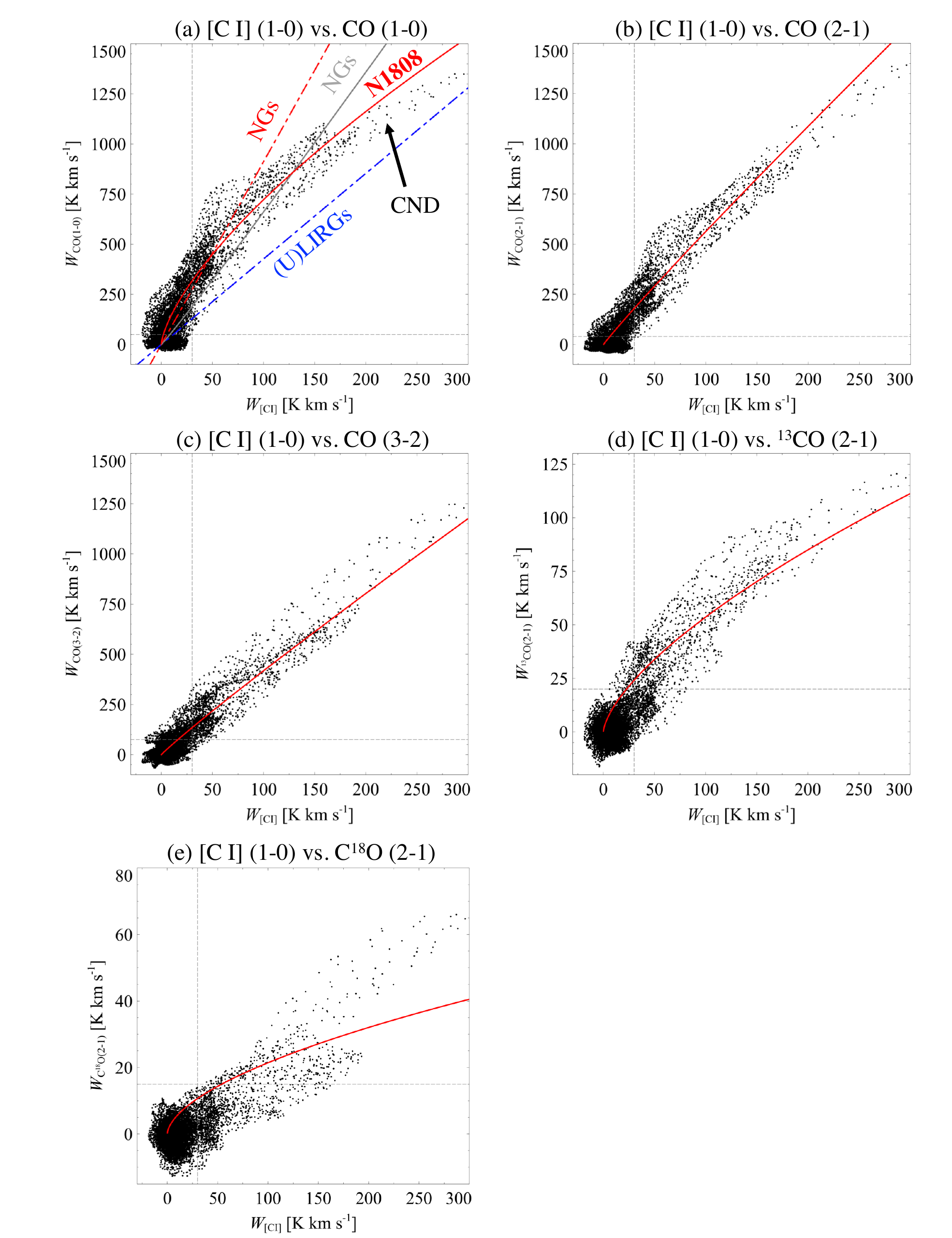}
 \caption{Integrated intensity (\(W\)) correlations. The solid red curve is a power law fit. The pixels with values above \(\approx5\sigma\) (gray dashed lines) were used for fitting. In panel (a), the gray curve is the fit for nearby galaxies (NGs), the dash-dotted red is a linear fit for nearby galaxies, and the dash-dotted blue is for (U)LIRGs and the Spiderweb Galaxy \citep{Jia19}. All data are smoothed to the resolution of CO (1--0) (\(\sim100\) pc); the pixel size (\(0\farcs4\)) is \(\sim1/5\) of the beam size (\(2\farcs666\times1\farcs480\)).\label{fig:cor}}
\end{figure*}

The comparison of all data points (pixels) is shown in Figure \ref{fig:cor}. We fitted the distribution in each panel by a power law \(W_\mathrm{CO}=aW_\mathrm{[CI]}^b\) using a least-squares method, and the resulting fitting parameters are listed in Table \ref{tab:fit}. Nearly linear fits are found only for CO (2--1) and CO (3--2), and the scatter is smallest for CO (2--1). Most importantly, the slope of the pixel distribution for \(W_\mathrm{CO(2-1)}\)--\(W_\mathrm{[CI]}\) in the CND is approximately the same as the one in the starburst disk; this is expected from the uniform intensity ratio in the central 1 kpc discussed in section \ref{sec:radpro}. On the other hand, CO (1--0), \(^{13}\)CO (2--1), and C\(^{18}\)O (2--1) exhibit significantly different slopes in the disk and the CND, which results in non-linear correlations. Figure \ref{fig:cor} shows that the discrepancy is particularly large for C\(^{18}\)O (2--1). The fit is dominated by low-intensity emission from the disk and largely changes slope in the CND; there is no single solution that can account for both regions. Interestingly, the relation \(T_\mathrm{b,CO(1-0)}\mathrm{-}T_\mathrm{b,[CI]}\) in Orion A presented in Figure 4 in \citet{Shi13} appears to be similar at a spatial scale of only 0.04 pc. The correlation between [\ion{C}{1}] and CO (1--0) at 100 pc resolution in NGC 1808 is a power of \(\sim0.7\). Both \citet{Shi13} and \citet{Ike02} found near-linear relations between \(^{13}\)CO (1--0) and [\ion{C}{1}] in Orion. Approximately linear correlations in Orion were also reported for \(^{13}\)CO (2--1) and [\ion{C}{1}] by \citet{Tau95}; see \citet{Kee97} for a review on the correlations in a number of Galactic molecular clouds. The relation between \(^{13}\)CO (2--1) and [\ion{C}{1}] in NGC 1808 is less linear, largely due to a different slope in the CND, and the scatter is large.

In Figure \ref{fig:cor}(a), we compare the results with the fits obtained by \citet{Jia19} for a sample of various galaxy types at low resolution. The gray line is the power law \(\log{L'}_\mathrm{CO(1-0)}=0.74+1.04\log{L'}_\mathrm{[CI]}\) for nearby galaxies at \(\sim1\) kpc resolution. We showed in Figure \ref{fig:lum} that the total luminosity of NGC 1808 is in excellent agreement with this fit. Also shown are linear fits \(\log{L'}_\mathrm{CO(1-0)}=0.96+\log{L'}_\mathrm{[CI]}\) for the same sample of nearby galaxies (red dash-dotted line) and \(\log{L'}_\mathrm{CO(1-0)}=0.63+\log{L'}_\mathrm{[CI]}\) (blue dash-dotted line) for (U)LIRGs and the Spiderweb Galaxy \citep{Jia17,Jia19}. Note that the change in slope that we observe at high resolution in NGC 1808 can be regarded as a combination of two regions with different physical conditions: 1 kpc disk, where the slope is comparable to the average for nearby galaxies, and the CND and other hot spots, where the slope is comparable to the (U)LIRGs. We conclude that the physical conditions of molecular gas play an important role in determining the [\ion{C}{1}] (1--0) -- CO (1--0) correlation, and that the correlation departures from linearity when different galactic environments are studied at a resolution higher than 1 kpc. It would be interesting to investigate the correlations at high resolution across entire galactic disks, beyond the central 1 kpc studied here.

\begin{table}
\begin{center}
\caption{Fitting Parameters for CO--[\ion{C}{1}] Correlations}\label{tab:fit}
\begin{tabular}{lccc}
\tableline\tableline
Line & \(a\) & \(b\) & \(c\) \\
\tableline
CO (1--0) & \(1.488\pm0.019\) & \(0.686\pm0.010\) & \(0.85\) \\
CO (2--1) & \(0.854\pm0.024\) & \(0.948\pm0.014\) & \(0.86\) \\
CO (3--2) & \(0.738\pm0.025\) & \(0.942\pm0.014\) & \(0.86\) \\
\(^{13}\)CO (2--1) & \(0.401\pm0.026\) & \(0.664\pm0.014\) & \(0.83\) \\
C\(^{18}\)O (2--1) & \(0.171\pm0.068\) & \(0.580\pm0.033\) & \(0.66\) \\
\tableline
\end{tabular}
\end{center}
\tablecomments{\(c\) is the correlation coefficient.}
\end{table}

\subsubsection{Does [\ion{C}{1}] (1--0) Trace Molecular Gas Mass?}

In section \ref{sec:radpro}, it was pointed out that \(r_{10}\) depends on radius. Similarly, \citet{Kri16} have shown that the [\ion{C}{1}]/CO (1--0) ratio is not uniform in the central starburst of NGC 253. If the distribution of \ion{C}{1} is well-mixed with that of H\(_2\) molecules inside molecular clouds, and the intensity of [\ion{C}{1}] (1--0) emission is proportional to the column density \(N_\mathrm{CI}\propto N_\mathrm{H_2}\) (optically thin case), this would imply that CO (2--1) emission, that arises predominantly from the cloud envelopes, is also proportional to \(N_\mathrm{H_2}\). However, the CO-to-H\(_2\) conversion factor \(\alpha_\mathrm{CO}\) based on CO (1--0) is thought to be lower than the standard Galactic disk value in starbursts, because \(\alpha_\mathrm{CO}\propto \sqrt{n_\mathrm{H_2}}/T_\mathrm{k}\) and clouds may not be virialized \citep{Bol13}. The relatively high CO (2--1)/(1--0) ratio of \(\sim1\) and the uniform \(r_{21}\) then imply that the conversion factors based on [\ion{C}{1}] (1--0) or CO (2--1) should be lowered even more for the starburst region compared to the values applied to the outer disk. This result suggests that the CO (1--0) based conversion factor is likely superior to those based on [\ion{C}{1}] (1--0) and CO (2--1) when applied universally, regardless of galaxy type (see also \citealt{IRv15,Val18}).

On the other hand, the \(L'_\mathrm{CO(1-0)}\mathrm{-}L'_\mathrm{[CI]}\) correlation is nearly linear and tight when star-forming galaxies, such as local spirals, are observed at a resolution of 1 kpc \citep{Jia19}. This suggests that a large fraction of [\ion{C}{1}] (1--0) flux in such galaxies may originate from a relatively cold (\(T_\mathrm{k}\sim20\) K), low-density (\(n_\mathrm{H_2}\sim10^3\) cm\(^{-3}\)) disk, where the intensity ratios are comparable to the values in the Galactic disk (\(r_{10}\sim0.15\); \citealt{FBM99}). Figure \ref{fig:tbr} shows that \(r_{10}\sim0.15\) and \(r_{21}\sim0.20\) at radii larger than the starburst disk in NGC 1808. On the other hand, it was demonstrated in section \ref{sec:rad} and in \citet{Sal18} that the physical conditions in the CND are more extreme (higher gas temperature \(T_\mathrm{k}\sim40\mathrm{-}80\) K and density \(n_\mathrm{H_2}\sim10^{3\mathrm{-}4}\) cm\(^{-3}\)) compared to the conditions in typical molecular clouds far from star-forming regions (\(T_\mathrm{k}\sim10\mathrm{-}20\) K and \(n_\mathrm{H_2}\sim10^{2\mathrm{-}3}\) cm\(^{-3}\); e.g., \citealt{WWT97,Eva99}). Both high excitation due to physical conditions and high \ion{C}{1} abundance in the CND contribute to an enhanced [\ion{C}{1}]/CO(1--0) luminosity ratio that may overestimate the total H\(_2\) gas mass compared to that derived using \(\alpha_\mathrm{CO}\). Unless the total flux is dominated by warm and dense gas, such as the conditions in the CND of NGC 1808 and (U)LIRGs, the near-linear relation for nearby galaxies is expected to hold and a conversion factor for molecular gas may be established based on [\ion{C}{1}] (1--0) luminosity, in a similar manner that CO (2--1) is often used (e.g., \citealt{Ler13,San13}).

The tight correlations between [\ion{C}{1}] and optically thick CO (2--1) and CO (3--2) lines also support the scenario that, at least in the starburst region, [\ion{C}{1}] (1--0) emission may be arising predominantly from the outer layers of (clumpy) molecular clouds, in agreement with photodissociation region models \citep{Spa96,HT97}.

\subsection{Atomic Carbon in the Outflow}\label{sec:ofl}

Some recent studies have suggested that atomic carbon abundance can be enhanced in cosmic-ray dominated regions such as starburst nuclei and molecular outflows \citep{PTV04,PBZ18,Bis17}. For example, outflows detected in [\ion{C}{1}] are reported for NGC 253 (starburst-driven), NGC 613 (AGN-driven), and NGC 6240 \citep{Kri16,Miy18,Cic18}.

At a resolution of 30 pc, we could not identify a \ion{C}{1} outflow from the location of the AGN in NGC 1808 as the spectrum toward the core does not exhibit high-velocity components. This is consistent with the picture that the AGN feedback is weak and that the dust outflow is starburst-driven and generated at larger scales.

To search for \ion{C}{1} in the large-scale outflow, we analyzed the position-velocity space in directions that coincide with some of the prominent polar dust lanes observed as absorption in optical images and where CO (1--0) was detected. The investigated regions also exhibit extended emission of ionized gas and enhanced [\ion{N}{2}]/H\(\alpha\) intensity ratio suggesting large-scale shocks \citep{SB10}. To increase sensitivity and enable direct comparison, the [\ion{C}{1}] data were smoothed to the resolution of CO (1--0) (\(\sim100\) pc).

\begin{figure*}
 \centering
  \includegraphics[width=0.9\textwidth]{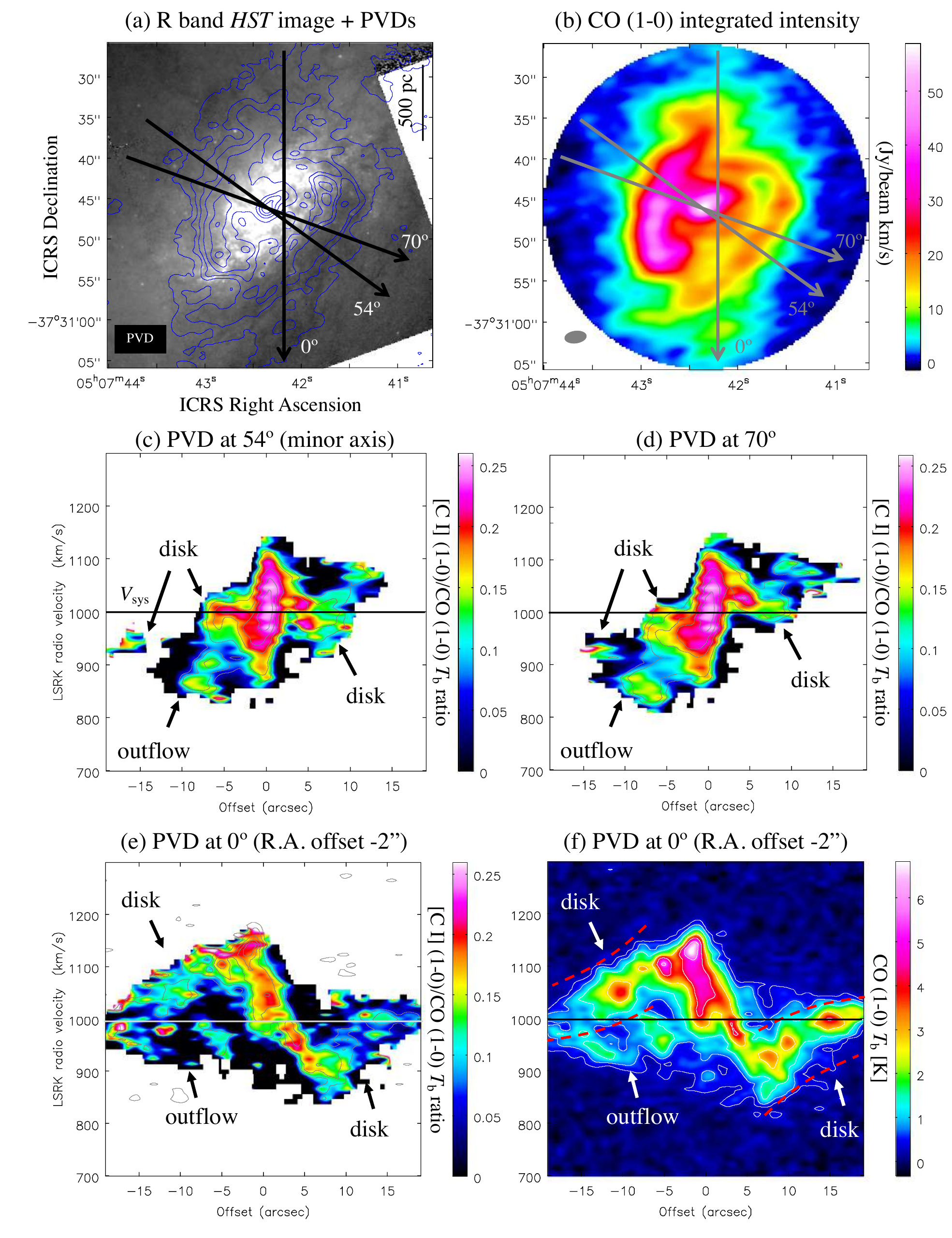}
 \caption{Position-velocity diagrams along three directions (\(PA=0\degr\), \(54\degr\), and \(70\degr\)) indicated in panels (a,b). (b) CO (1--0) integrated intensity. Panels (c-e) show [\ion{C}{1}] (1--0)/CO (1--0) intensity ratios (in K) where the contours are [\ion{C}{1}] (1--0) at \((0.1,0.2,0.4,0.6,0.8)\times S_\mathrm{max}\) [Jy beam\(^{-1}\)]. (f) CO (1--0) intensity, where the contours are \((0.05,0.1,0.2,0.4,0.6,0.8)\times 8.41\) K.\label{fig:pvds}}
\end{figure*}

The constructed position-velocity diagrams (PVDs), presented in Figure \ref{fig:pvds}, show that [\ion{C}{1}] emission is relatively weak and detected only toward the base of the outflow (central 1 kpc). Toward the minor galactic axis, shown in panel (c), [\ion{C}{1}] is detected in the outflow component with a [\ion{C}{1}] (1--0)/CO (1--0) intensity ratio of \(r_{10}\sim0.15\). Here, the outflow (marked by an arrow) is identified where the line-of-sight velocity is \(v\sim-100\) km s\(^{-1}\) relative to the disk component. There is also a weak component on the opposite side (offset \(+5\arcsec\), relative velocity up to \(v\sim+100\) km s\(^{-1}\)) detected only in CO. For an inclination of \(57\arcdeg\), the average velocity of the outflow perpendicular to the galactic disk is \(v_\mathrm{out}\sim180\) km s\(^{-1}\). The direction in Figure \ref{fig:pvds}(e) is offset \(\Delta(\mathrm{R.A.})-2\arcsec\) from the center in order to coincide with a dust lane that emerges from \(\sim5\arcsec\) northward (see also \citealt{Phi93}). Here, the CO (1--0) line width is 150 km s\(^{-1}\) and line splitting is present (line-of-sight velocity component at \(v\sim-100\) km s\(^{-1}\) relative to the disk component). This feature is likely associated with the extraplanar dust lanes. Here, atomic carbon is detected mostly in the disk behind the dust lane.

The PVDs in Figure \ref{fig:pvds} indicate that the intensity ratio is \(r_{10}\lesssim0.15\) in most outflow components. This is lower than the value in the CND (0.22) and comparable to or lower than in the starburst disk where the observed ratio is 0.15--0.20. The values are similar to those observed toward the central region of the superwind galaxy M82 at a resolution of 0.7 kpc \citep{Jia19}. This result can be explained by two possibilities: low \ion{C}{1} abundance in the outflow, or low density. For example, RADEX LVG calculations for \(n_\mathrm{H_2}=10^2\) cm\(^{-3}\), \(T_\mathrm{k}=50\) K, \(N/\Delta V=1.0\times10^{16}~\mathrm{cm^{-2}~(km~s^{-1})^{-1}}\), and \(N_\mathrm{CO}=N_\mathrm{CI}\) yield a ratio of \(\sim0.14\) even though the abundances of \ion{C}{1} and CO are the same. This is consistent with results from \citet{Sal18}, who estimated that the beam-averaged gas density in the outflow is of the order \(n_\mathrm{H_2}\sim10^{2\mathrm{-}3}\) cm\(^{-3}\).

Following the analysis in \citet{Sal16}, we estimate that the noncircular motions due to outflows comprise less than 10\% of the total [\ion{C}{1}] flux in the central region, which is \(M_\mathrm{out}\sim1\times10^4~M_\sun\). With an average outflow velocity of \(v_\mathrm{out}\sim180\) km s\(^{-1}\) perpendicular to the galactic disk, the upper limit of the kinetic energy of the atomic carbon outflow is \(E_\mathrm{k}\sim M_\mathrm{out}v_\mathrm{out}^2/2\sim3\times10^{51}~\mathrm{erg}\). The energy is five orders of magnitude smaller than the estimated energy output from supernova explosions \citep{Sal16}.

\section{Summary}\label{sec:sum}

We have reported comprehensive ALMA observations of [\ion{C}{1}] (1--0), low-\(J\) CO lines, and dense gas tracers toward the starburst galaxy NGC 1808 at a resolution of 30--50 pc. The main findings are summarized below.

\begin{enumerate}

\item{The first high-resolution images of [\ion{C}{1}] (1--0), CO, \(^{13}\)CO, C\(^{18}\)O (2--1), CS (5--4), and HNCO (10--9) were acquired toward the central radius 1 kpc region. Neutral atomic carbon [\ion{C}{1}] (1--0) was detected toward the starburst disk at a resolution of \(0\farcs6\) (30 pc) with distribution and kinematics similar to those of CO (2--1).}

\item{Non-LTE radiative transfer analysis indicates the presence of warm (\(T_\mathrm{k}=40\mathrm{-}80\) K) and dense (\(n_\mathrm{H_2}=10^{3\mathrm{-}4}\) cm\(^{-3}\)) molecular gas in the CND with a high atomic carbon column density of \(N_\mathrm{CI}\sim0.5N_\mathrm{CO}\sim3\times10^{18}\) cm\(^{-2}\). The \ion{C}{1}/H\(_2\) abundance in the central 1 kpc is \(3\mathrm{-}7\times10^{-5}\), similar to the values reported for starburst and luminous infrared galaxies.}

\item{The line intensity ratios of [\ion{C}{1}] (1--0) and five low-\(J\) CO lines were studied for the first time for an external galaxy at \(<100\) pc resolution. We found that the [\ion{C}{1}](1--0)/CO(1--0) and [\ion{C}{1}](1--0)/CO(3--2) intensity ratios exhibit negative and positive azimuthally-averaged gradients, respectively. By contrast, [\ion{C}{1}](1--0)/CO(2--1) is uniform in the central 1 kpc. This is explained by excitation and optical depth effects: the critical density and excitation temperature of CO (2--1) are similar to those of [\ion{C}{1}] (1--0). The intensities of \(^{13}\)CO and C\(^{18}\)O (2--1) relative to [\ion{C}{1}] vary by \(\sim30\%\) in the central \(R\lesssim400\) pc.}

\item{We studied the correlations between [\ion{C}{1}] and CO integrated intensities. Approximately linear correlations are found for CO (2--1) and CO (3--2), whereas the correlation with CO (1--0) is a power law \(W_\mathrm{CO(1-0)}\propto W_\mathrm{[CI]}^{0.7}\). The correlation with \(^{13}\)CO (2--1) is similarly \(W_\mathrm{^{13}CO(1-0)}\propto W_\mathrm{[CI]}^{0.7}\), while that with C\(^{18}\)O (2--1) could not be fitted with a single power law. These results suggest that physical conditions strongly affect the observed intensities and caution is needed when [\ion{C}{1}] (1--0) luminosity is used as a tracer of molecular gas mass in resolved galaxies with starburst regions. Since the correlation between CO (1--0) and [\ion{C}{1}] (1--0) in nearby galaxies is tight and nearly linear on kpc scale, a universal [\ion{C}{1}]-based conversion factor may still be applied if the measured [\ion{C}{1}] flux is not dominated by extreme physical conditions, such as the CND in NGC 1808 and (U)LIRGs. The excellent correlations between [\ion{C}{1}] (1--0) and optically thick CO (2--1) and CO (3--2) lines support the PDR scenario where [\ion{C}{1}] (1--0) emission arises predominantly from the outer layers of (clumpy) molecular clouds, at least in the starburst environment.}

\item{The [\ion{C}{1}]/CO (1--0) intensity ratio is \(\lesssim0.15\) toward the base of the starburst-driven outflow that emerges from the central 1 kpc region, comparable or less than in the starburst disk. The low ratio is possibly a consequence of low gas density (\(n_\mathrm{H_2}\lesssim10^{2\mathrm{-}3}\) cm\(^{-3}\)) averaged in an aperture of 100 pc. The upper limits of the mass and kinetic energy of the atomic carbon outflow are \(M_\mathrm{out}\sim1\times10^4~M_\sun\) and \(E_\mathrm{k}\sim3\times10^{51}~\mathrm{erg}\), respectively.}

\end{enumerate}

\acknowledgments

The authors thank the anonymous referee for many detailed comments and suggestions. This paper makes use of the following ALMA data: ADS/JAO.ALMA \#2012.1.01004.S, \#2013.1.00911.S, \#2016.1.00296.S, and \#2017.1.00984.S. ALMA is a partnership of ESO (representing its member states), NSF (USA) and NINS (Japan), together with NRC (Canada), MOST and ASIAA (Taiwan), and KASI (Republic of Korea), in cooperation with the Republic of Chile. The Joint ALMA Observatory is operated by ESO, AUI/NRAO and NAOJ. Based on observations made with the NASA/ESA \emph{Hubble Space Telescope} and obtained from the Hubble Legacy Archive, which is a collaboration between the Space Telescope Science Institute (STScI/NASA), the Space Telescope European Coordinating Facility (ST-ECF/ESA), and the Canadian Astronomy Data Centre (CADC/NRC/CSA). This research has made use of the NASA/IPAC Extragalactic Database (NED), which is operated by the Jet Propulsion Laboratory, California Institute of Technology, under contract with the National Aeronautics and Space Administration.

\clearpage

\appendix

\section{Column density of \ion{C}{1} in LTE}\label{app:A}

We consider the fine structure of \ion{C}{1} in ground state \(^3\mathrm{P}\). The absorption coefficient for a two-level (\(J=0,1\)) system in LTE can be expressed (e.g., \emph{Tools of Radio Astronomy}; \citealt{WRH13}) as

\begin{equation}
\kappa=\frac{c^2}{8\pi\nu^2}\frac{g_1}{g_0}n_0A_{10}\left(1-\mathrm{e}^{-h\nu/{kT_\mathrm{ex}}}\right)\varphi(\nu),
\end{equation}
where \(g_J=2J+1\) is the statistical weight, \(n_0\) is the density in level \({^3}\mathrm{P}_0\), \(A_{10}\) is the Einstein coefficient for spontaneous emission \(J=1\rightarrow0\), and \(\varphi(\nu)\) is the line profile defined as \(\int_0^\infty\varphi(\nu)\,d\nu=1\). From the Boltzmann distribution and approximation \(\varphi(\nu)\approx\frac{1}{\Delta \nu}\approx\frac{c}{\nu\Delta V}\), where \(\Delta V\) is the velocity width,

\begin{equation}
\kappa=\frac{c^2}{8\pi\nu^2}n_1A_{10}\left(\mathrm{e}^{h\nu/{kT_\mathrm{ex}}}-1\right)\varphi(\nu)=\frac{c^3}{8\pi\nu^3}n_1A_{10}\frac{1}{\Delta V}\left(\mathrm{e}^{h\nu/{kT_\mathrm{ex}}}-1\right).
\end{equation}

The optical depth of a line is the absorption coefficient integrated over the line of sight, \(\tau\equiv\int\kappa\,ds\). Defining a column density, \(N\equiv\int n\,ds\), we get

\begin{equation}\label{eq:tau}
\tau=\frac{c^3}{8\pi\nu^3}N_1A_{10}\frac{1}{\Delta V}\left(\mathrm{e}^{h\nu/{kT_\mathrm{ex}}}-1\right).
\end{equation}
In general, the column density in level \(i\) relative to level \(J\) is

\begin{equation}
\frac{N_i}{N_J}=\frac{g_i}{g_J}\mathrm{e}^{-(E_i-E_J)/{kT_\mathrm{ex}}},
\end{equation}
and the total column density of \ion{C}{1} is \(N_\mathrm{CI}=\sum_{i=0}^2 N_i\). Then,

\begin{equation}
N_\mathrm{CI}=\frac{N_J}{g_J}\mathrm{e}^{E_J/{kT_\mathrm{ex}}}\sum_{i=0}^2 g_i\mathrm{e}^{-E_i/{kT_\mathrm{ex}}}
\end{equation}
and

\begin{equation}
\frac{N_1}{N_\mathrm{CI}}=\frac{g_1}{Q}\mathrm{e}^{-E_1/{kT_\mathrm{ex}}},
\end{equation}
where \(E_1=h\nu\) is the energy of the \({^3}\mathrm{P}_1\) level, and \(Q\) is the partition function defined as \(Q=\sum_{i=0}^2 g_i\mathrm{e}^{-{E_i}/{kT_\mathrm{ex}}}\). In this case, there are three levels (\(E_0=0\) is the ground state), hence \(Q=1+3\mathrm{e}^{-E_1/{kT_\mathrm{ex}}}+5\mathrm{e}^{-E_2/{kT_\mathrm{ex}}}\). The energies of the levels are \(E_1/k=23.6\) K and \(E_2/k=62.5\) K.

Using equation \ref{eq:tau}, the total column density of \ion{C}{1} becomes

\begin{equation}
N_\mathrm{CI}=\frac{8\pi\nu^3}{c^3A_{10}}\tau\Delta V\frac{Q}{g_1}\frac{\mathrm{e}^{E_1/{kT_\mathrm{ex}}}}{\mathrm{e}^{E_1/{kT_\mathrm{ex}}}-1}.
\end{equation}
With a definition of radiation temperature, \(J(T_\mathrm{ex})=(h\nu/k)(\mathrm{e}^{E_1/kT_\mathrm{ex}}-1)^{-1}\), we get

\begin{equation}
N_\mathrm{CI}=\frac{8\pi k\nu^2}{hc^3A_{10}}J(T_\mathrm{ex})\frac{Q}{g_1}\mathrm{e}^{E_1/{kT_\mathrm{ex}}}\tau\Delta V,
\end{equation}
and using \(T_\mathrm{b}=[J(T_\mathrm{ex})-J(T_\mathrm{cmb})](1-\mathrm{e}^{-\tau})\approx J(T_\mathrm{ex})(1-\mathrm{e}^{-\tau})\) and \(W=T_\mathrm{b}\Delta V\), the expression can be written as

\begin{equation}\label{cdci}
N_\mathrm{CI}=\frac{8\pi k\nu^2}{hc^3A_{10}}\frac{Q}{g_1}\mathrm{e}^{E_1/{kT_\mathrm{ex}}}\frac{\tau}{1-\mathrm{e}^{-\tau}}W.
\end{equation}

\section{Ratio maps}\label{app:B}

\begin{figure*}
 \centering
  \includegraphics[width=0.85\textwidth]{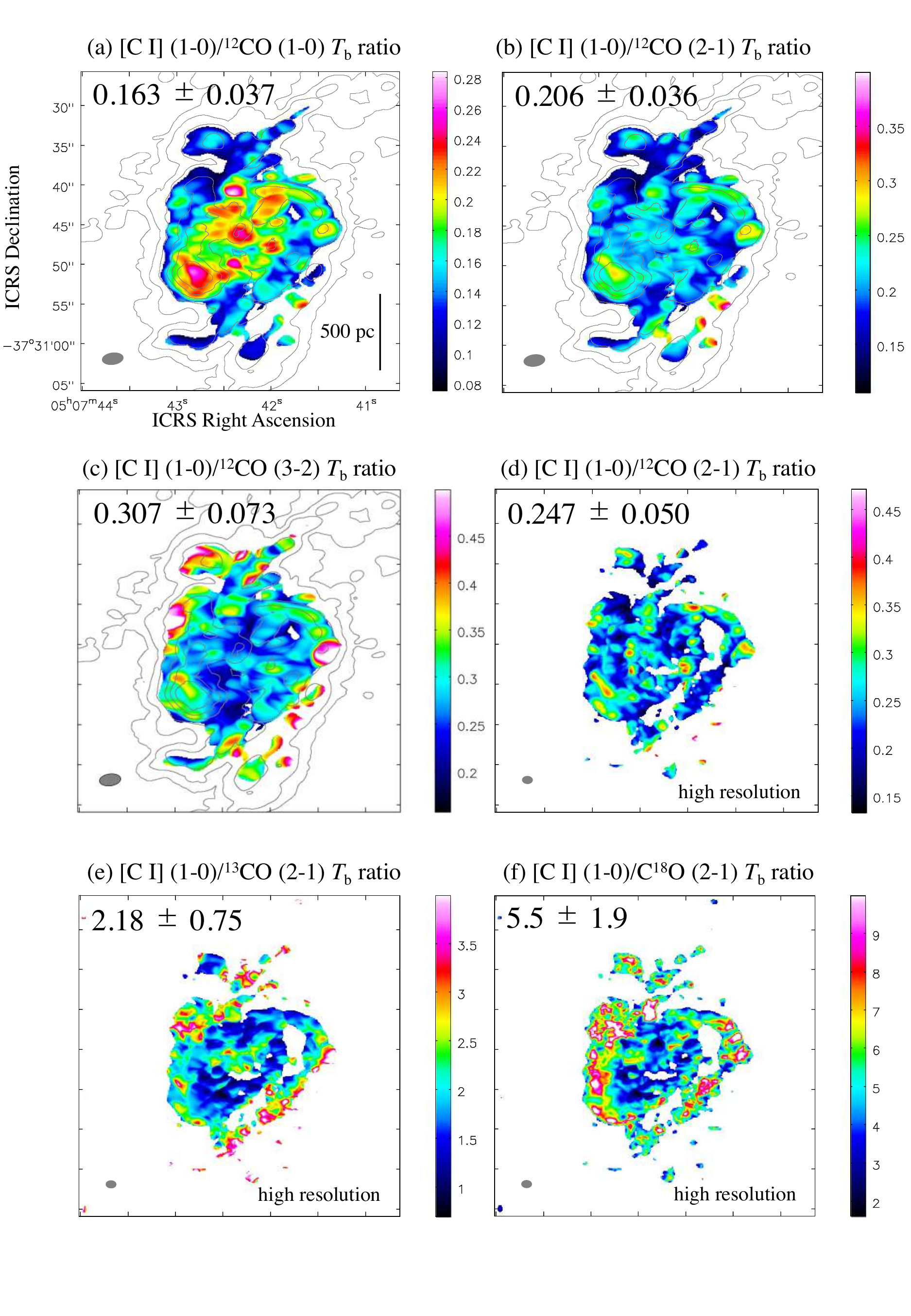}
 \caption{Peak brightness temperature (\(T_\mathrm{b}\)) ratios of [\ion{C}{1}] and CO lines. The contours in panels (a-c) are the CO (1--0) \(T_\mathrm{b}\) plotted at \((0.05,0.1,0.2,0.4,0.6,0.8)\times13.95~\mathrm{K}\) (maximum). The ratio images are clipped below \(2\sigma\) of the [\ion{C}{1}] image, where \(1\sigma=0.16~\mathrm{K}\) at the resolution of \(2\farcs666\times1\farcs480\) in panels (a-c) and \(1\sigma=0.30~\mathrm{K}\) at the resolution of \(1\farcs280\times0\farcs943\) in panels (d-f). Mean ratios and standard deviations are shown at the top left corner.\label{fig:tbr1}}
\end{figure*}

In Figure \ref{fig:tbr1}, we show the brightness temperature (\(T_\mathrm{b}\)) ratio maps of all tracers. Note that \(T_\mathrm{b}\) here was calculated using the Rayleigh-Jeans formula, so that the \(T_\mathrm{b}\) ratio of two lines is proportional to the ratio of their fluxes \(S\) as \(T_\mathrm{b2}/T_\mathrm{b1}\propto(\nu_1/\nu_2)^2 S_2/S_1\). This is equivalent to the main beam brightness temperature used in single dish observations and differs from the brightness temperature in the Planck law at these frequencies, because the Rayleigh-Jeans approximation is not valid. According to the Rayleigh-Jeans law, the brightness is given by

\begin{equation}\label{eq:planck}
B_\nu=\frac{2k\nu^2}{c^2}T_\mathrm{b}.
\end{equation}

\begin{figure*}
 \centering
  \includegraphics[width=0.85\textwidth]{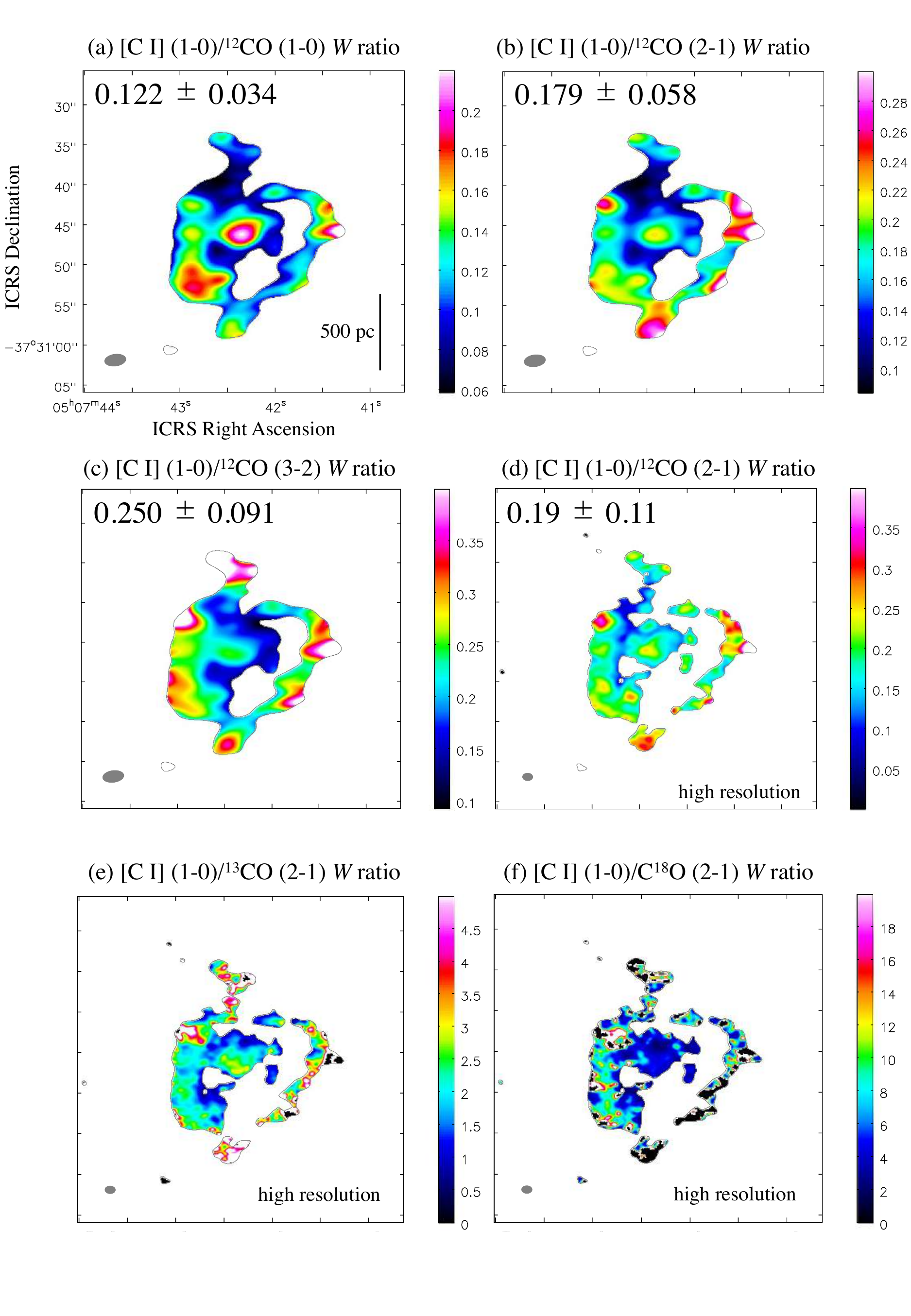}
 \caption{Integrated intensity (\(W\)) ratios of [\ion{C}{1}] and CO lines. The ratio images are clipped below \(4\sigma\) of the [\ion{C}{1}] (1--0) integrated intensity, where \(1\sigma=8\) K km s\(^{-1}\) at the resolution of \(2\farcs666\times1\farcs480\) in panels (a-c) and \(1\sigma=11\) K km s\(^{-1}\) at the resolution of \(1\farcs280\times0\farcs943\) in panels (d-f). Mean ratios and standard deviations are shown at the top left corner except in panels (e,f) because of poor signal-to-noise ratio in the outer regions.\label{fig:wr}}
\end{figure*}

Figure \ref{fig:wr} shows the integrated intensity (\(W\)) ratio maps. \(W\) can be related to the total brightness by

\begin{equation}
\left(\frac{B}{\mathrm{erg~s^{-1}~cm^{-2}~sr^{-1}}}\right)=1.025\times10^{-15}\left(\frac{\nu}{\mathrm{GHz}}\right)^3\left(\frac{W}{\mathrm{K~km~s^{-1}}}\right),
\end{equation}
and the integrated intensity ratio can then be expressed as brightness ratio by multiplying \((\nu_\mathrm{[CI]}/\nu_\mathrm{CO})^3\).

\end{document}